\documentclass{aa}
\newcommand{\msun}{$M_\odot$}
\newcommand{\vmic}{$\xi_t$}

\newcommand{\ittt}{p$T_{\rm eff}$}
\newcommand{\iggg}{p$\log g$}

\newcommand{\fxx}{^{\rm F}}

\newcommand{\lamboo}{$\lambda$~Boo}

\newcommand{\dss}{$\delta$~Scuti}
\newcommand{\rms}{{\em rms}}

\newcommand{\bfs}{bf.\ $\delta$~Scu}
\newcommand{\bfss}{bona fide $\delta$~Scuti}
\newcommand{\bon}{{bf.\ $\gamma$~Dor}}
\newcommand{\bonn}{{bona fide $\gamma$~Dor}}

\newcommand{\reff}{reference}

\newcommand{\ione}{{\,\sc i}}
\newcommand{\itwo}{{\,\sc ii}}

\newcommand{\scii}{{\,\sc ii}}
\newcommand{\gamdor}{$\gamma$~Dor}
\newcommand{\gamdorr}{$\gamma$~Doradus}

\newcommand{\eeee}{El.}
\newcommand{\atlasni}{ATLAS9} 
\newcommand{\vwa}{{\sc VWA}}

\newcommand{\cons}{constant}
\newcommand{\can}{cand.\ $\gamma$~Dor}
\newcommand{\candd}{candidate $\gamma$~Dor}
\newcommand{\templogg}{{\sc TEMPLOGG}}
\newcommand{\synth}{{\sc SYNTH}}
\newcommand{\simbad}{{\sc SIMBAD}}
\newcommand{\vald}{{\sc VALD}}
\newcommand{\hipp}{{\sc HIPPARCOS}}
\newcommand{\str}{Str\"omgren}

\newcommand{\topp}{{\em top}}
\newcommand{\midd}{{\em middle}}
\newcommand{\bott}{{\em bottom}}
\newcommand{\lee}{{\em left}}
\newcommand{\rii}{{\em right}}
\newcommand{\panel}{{\em panel}}
\newcommand{\panels}{{\em panels}}

\newcommand{\eg}{e.g.}
\newcommand{\cf}{cf.}
\newcommand{\ie}{i.e.}

\newcommand{\chitwo}{$\chi^2$}

\newcommand{\paper}{Paper~I}

\newcommand{\loggf}{$\log gf$}

\newcommand{\teff}{$T_{\rm eff}$}

\newcommand{\logg}{$\log g$}

\newcommand{\feh}{[Fe/H]}
\newcommand{\meh}{[M/H]}
\newcommand{\cah}{[Ca/H]}
\newcommand{\vsini}{$v \sin i$}
\newcommand{\kms}{km\,s$^{-1}$}

\usepackage{txfonts}
\usepackage{graphicx}
 
\usepackage{natbib}
\bibpunct{(}{)}{;}{a}{}{,} 

\begin{document}
   \title{A spectroscopic study of southern (candidate) $\gamma$~Doradus stars.}
   \subtitle{II. Detailed abundance analysis and fundamental parameters}
   \author{H.\ Bruntt
          \inst{1,2}
          \and
          P.\ De~Cat 
          \inst{3,4}
          \and
          C.\ Aerts
          \inst{4,5}
          }
   \offprints{H. Bruntt}
   \institute{Niels Bohr Institute, University of Copenhagen,
              Juliane Maries Vej 30, DK-2100 Copenhagen \O, Denmark
         \and
             {School of Physics A28, University of Sydney, 2006 NSW, Australia}
              \email{bruntt@physics.usyd.edu.au}
	 \and
              Royal Observatory of Belgium, Ringlaan 3, B-1180 Brussel, Belgium
              \email{peter@oma.be}
         \and
              Katholieke Universiteit Leuven, Celestijnenlaan 200B, B-3001 Leuven, Belgium
              \email{conny@ster.kuleuven.ac.be}
          \and
        Department of Astrophysics, Radboud University Nijmegen, 6500 GL Nijmegen, the Netherlands
             }
   \date{Received xxx-xxx 2007; accepted yyy-yyy 2007}
  \abstract
{The \gamdorr\ stars are a recent class of variable main sequence F-type stars
located on the red edge of the Cepheid instability strip. 
They pulsate in gravity modes, and this makes them particularly interesting
for detailed asteroseismic analysis, which can provide fundamental knowledge 
of properties near the convective cores of intermediate-mass main sequence stars.}  
{To improve current understanding of \gamdor\ stars through theoretical modelling,
additional constraints are needed. Our aim is to estimate
the fundamental atmospheric parameters and determine the chemical composition of these stars.
Detailed analyses of single stars have previously suggested links to Am and \lamboo\ stars, so 
we wish to explore this interesting connection between chemical peculiarity and pulsation.} 
{We have analysed a sample of \gamdor\ stars for the first time, including nine bona fide and three 
candidate members of the class. We determined the fundamental atmospheric parameters and 
compared the abundance pattern with other A-type stars. 
We used the semi-automatic software package \vwa\ for the analysis. 
This code relies on the calculation of synthetic spectra and thus takes line-blending into account. 
This is important because of the fast rotation in 
some of the sample stars, and we made a thorough analysis of how \vwa\ performs when increasing \vsini. 
We obtained good results in agreement with previously
derived fundamental parameters and abundances in a few selected
reference stars with properties similar to the \gamdor\ stars.
}  
{We find that the abundance pattern in the \gamdor\ stars is not distinct from 
the constant A- and F-type stars we analysed.} 
{} 


   \maketitle
%

   \begin{table*}
      \caption[]{Fundamental atmospheric parameters of the target stars as determined from photometric indices and parallaxes.
         \label{tab:fund}}
\centering                          
\begin{tiny}
\begin{tabular}{r|rr|r@{}r|lcc|c}
\hline\hline

 & \multicolumn{2}{c|}{\paper} & \multicolumn{2}{c|}{2MASS} & \multicolumn{3}{c|}{\str} & \multicolumn{1}{c}{\hipp}\\
\hline

       & \multicolumn{1}{c}{Variability} & \multicolumn{1}{c|}{\vsini} &  ($V$   &$-\,K$)      & ($b-y$)    &  $m_1$   &  $c_1$  & $\pi$, \teff, $M$ \& $V$ \\     
    HD & \multicolumn{1}{c}{     type  } & \multicolumn{1}{c|}{[\kms]} & \teff\ \,  & [K]  & \teff\ [K] &  \feh\   & \logg\  &  \logg      \\ \hline
		          
    7455 &       \cons &  3   &  $\bf6400\,\pm$&$  90$ &  $6070\pm250    $ & ${\bf-0.17}\pm0.10$ &  $ 4.33\pm0.20$ &  ${\bf3.94}\pm0.20  $     \\
   12901 &       \bon  & 64   &  $\bf6950\,\pm$&$  90$ &  $7200\pm250    $ & ${\bf-0.33}\pm0.10$ &  $ 4.39\pm0.20$ &  ${\bf4.07}\pm0.13  $     \\
   14940 &       \bon  & 39   &  $\bf7090\,\pm$&$ 100$ &  $7200\pm250    $ & ${\bf-0.23}\pm0.10$ &  $ 4.31\pm0.20$ &  ${\bf4.25}\pm0.12  $     \\
   22001 &       \cons & 13   &  $\bf7130\,\pm$&$ 550$ &  $6690\pm250    $ & ${\bf-0.07}\pm0.10$ &  $ 4.23\pm0.20$ &  ${\bf4.40}\pm0.19  $     \\
   26298 &       \can  & 50   &  $\bf6780\,\pm$&$  90$ &  $6730\pm250    $ & ${\bf-0.36}\pm0.10$ &  $ 4.12\pm0.20$ &  ${\bf3.95}\pm0.21  $     \\
   27290 &       \bon  & 54   &  $\bf7310\,\pm$&$ 500$ &  $7200\pm250    $ & ${\bf-0.01}\pm0.10$ &  $ 4.23\pm0.20$ &  ${\bf4.29}\pm0.18  $     \\
   27604 &       \cons & 70   &  $\bf6320\,\pm$&$  80$ &  $6450\pm250    $ & ${\bf+0.09}\pm0.10$ &  $ 3.80\pm0.20$ &  ${\bf3.65}\pm0.11  $     \\
   33262 &       \cons & 14   &  $\bf6060\,\pm$&$ 500$ &  $6130\pm250    $ & ${\bf-0.21}\pm0.10$ &  $ 4.58\pm0.20$ &  ${\bf4.62}\pm0.20  $     \\
   40745 &       \bon  & 37   &  $\bf6900\,\pm$&$ 100$ &  $6950\pm250    $ & ${\bf+0.08}\pm0.10$ &  $ 3.91\pm0.20$ &  ${\bf4.05}\pm0.12  $     \\
   48501 &       \bon  & 40   &  $\bf7240\,\pm$&$ 100$ &  $6980\pm250    $ & ${\bf-0.12}\pm0.10$ &  $ 3.92\pm0.20$ &  ${\bf4.28}\pm0.12  $     \\
   65526 &       \bon  & 53   &  $\bf7170\,\pm$&$ 110$ &  $   -          $ & $ -           $     &  $-           $ &  ${\bf4.40}\pm0.13  $     \\
   85964 &       \cons & 69   &  $\bf6600\,\pm$&$  90$ &  $6790\pm250    $ & ${\bf-0.03}\pm0.10$ &  $ 4.09\pm0.20$ &  ${\bf4.14}\pm0.13  $     \\
  110379 &       \can  & 24   &  $   5450\,\pm$&$ 420$ &  $6860\pm250    $ & ${\bf-0.17}\pm0.10$ &  $ 4.33\pm0.20$ &  $    4.39 \pm0.14  $     \\
  125081 &       \bfs  & 14   &  $\bf6380\,\pm$&$  90$ &  $6850\pm250    $ & ${\bf+0.54}\pm0.10$ &  $ 3.69\pm0.20$ &  ${\bf3.44}\pm0.20  $     \\
  126516 &       \can  &  4   &  $\bf6330\,\pm$&$  90$ &  $6630\pm250    $ & ${\bf-0.09}\pm0.10$ &  $ 4.37\pm0.20$ &  ${\bf4.17}\pm0.20  $     \\
  135825 &       \bon  & 38   &  $\bf7050\,\pm$&$  90$ &  $7230\pm250    $ & ${\bf-0.09}\pm0.10$ &  $ 4.30\pm0.20$ &  ${\bf4.39}\pm0.13  $     \\
  167858 &       \bon  & 13   &  $\bf7130\,\pm$&$ 100$ &  $7160\pm250    $ & ${\bf-0.12}\pm0.10$ &  $ 4.14\pm0.20$ &  ${\bf4.23}\pm0.12  $     \\
  218225 &       \bon  & 60   &  $\bf6920\,\pm$&$ 100$ &  $   -          $ & $ -  $              &  $-           $ &  ${\bf4.31}\pm0.21  $     \\ \hline

\multicolumn{9}{l}{Adopted values for the initial atmosphere models are printed in {\bf bold}.}

\end{tabular}
\end{tiny}
\end{table*}


\section{Introduction}

The members of the \gamdor\ class of variable stars are found near the
main sequence on the cool edge of the Cepheid instability strip with spectral types A7--F5.
They thus share properties with \dss\ star variables,
but the \gamdor\ periods are an order of magnitude longer, indicative of $g$~mode pulsation.
The \gamdor\ phenomenon was first identified by \cite{balona94}, 
and \cite{kris95} presented the first list of six candidates.
\cite{henry07} presents a list of 66 \gamdor\ stars, 
and the group continues to grow as new members are discovered both
among field stars \citep{mathias04, henry05, decat06} and in open clusters \citep{arentoft07}. 
Only a few stars show pulsations characteristic of \gamdor\ and \dss\ stars simultaneously \citep{henry05,rowe06,king07}.

The \gamdor\ stars have given new hope for a deeper understanding 
of main sequence stars with masses around 2\,\msun\ through asteroseismic analyses.
Several \dss\ stars have been studied extensively through both photometry and
spectroscopy, and dozens of individual modes are now known in a few field stars \citep{breger05} and also 
in members of open clusters \citep{bruntt07}. 
Whilst observational work has been very successful, comparison with theoretical models 
has so far not been able to provide a fully adequate description 
of all the observations (see \citealt{zima06} for recent developments). 
While the driving in \dss\ stars is well understood in terms of 
the opacity or $\kappa$ mechanism, 
the link to predicting observed mode amplitudes is weak.
Furthermore, theoretical models of \dss\ stars show that even moderate rotation 
leads to significant shifts in the mode frequencies \citep{suarez06a,suarez06b},
which complicates the confrontation of observations and models.

The theoretical framework for interpreting the observed pulsation in \gamdor\
stars is well under way. Pulsations are thought to be 
driven by a flux blocking mechanism near the base of their convective envelopes \citep{dupret04, dupret06}. 
\cite{moya05} investigated a method of constraining the models of \gamdor\ using
frequency ratios. 
This method has been attempted on individual \gamdor\ 
stars \citep{moya05,rodriguez06}. 
While the method is indeed very useful for providing constraints,
no unique models that fit all the observations were found.
An improvement would be to better constrain the fundamental
atmospheric parameters including metallicity. The star studied by
\cite{moya05}, HD~12901, is included in our sample.

In the current study we carry out a detailed abundance analysis
of a sample of \gamdor\ stars described by (\citealt{decat06}; hereafter \paper)
Thus, the current work is the second part of our detailed spectroscopic
analysis of a sample of southern candidate \gamdor\ stars. 
In \paper\ we made a detailed analysis of the
spectra to study binarity and the pulsation properties. 
We identified 10 new bona fide   
\gamdor\ stars of which 40\% are binary stars.

Detailed abundance analyses of \gamdor\ stars have only been done
for a few individual stars. 
\cite{bruntt02} analysed the \gamdor\ star HD~49434 
and found a metallicity slightly below solar,
but the analysis was hampered by the high \vsini\,$=85$\,\kms.
\cite{sadakane06} analysed
HD~218396 and found solar abundance of C and O (but not S) and abundances
of iron peak elements of $-0.5$~dex, thus suggesting a \lamboo\ nature for this star. 
\cite{henry05} found evidence that the \gamdor\ star HD~8801 is an Am star based on the
strength of the Ca~K line.
Our aim is to shed light on these intriguing links that have been
suggested between the \gamdor\ variables and
the chemically peculiar \lamboo\ and Am-type stars \citep{gray99,sadakane06}.

\section{Observations and selection of targets}

We have obtained high-resolution spectra with the \'echelle spectrograph
CORALIE attached to the 1.2-m Euler telescope (La Silla, Chile) for a sample
of 37 known and candidate $\gamma$\,Dor stars. For the details of the
observations and the data reduction, we refer to \paper. CORALIE covers
the 3880--6810\,\AA\ region in 68 orders with a spectral resolution of 50\,000.
The typical S/N in the spectra is 100--150. For the abundance analysis, we
selected the spectrum with the highest S/N. The wavelength calibrated
spectra were rebinned to a step size of $\simeq0.02$\,\AA. 
Each order was normalised by fitting low-order polynomials to 
continuum windows identified in a synthetic spectrum. 
The orders were then merged to a single spectrum 
while making sure the overlapping orders agreed. 

Stars with projected rotational velocities $v \sin i>70$\,\kms\ have 
not been analysed due to two reasons: only very few unblended lines are
available and we found that incorrect normalization of the spectra 
would introduce large systematic errors \citep{erspamer03}.
We also did not analyse the double-lined spectroscopic binaries
from \paper.

%
   \begin{figure*}
   \centering
   \includegraphics[width=17.6cm]{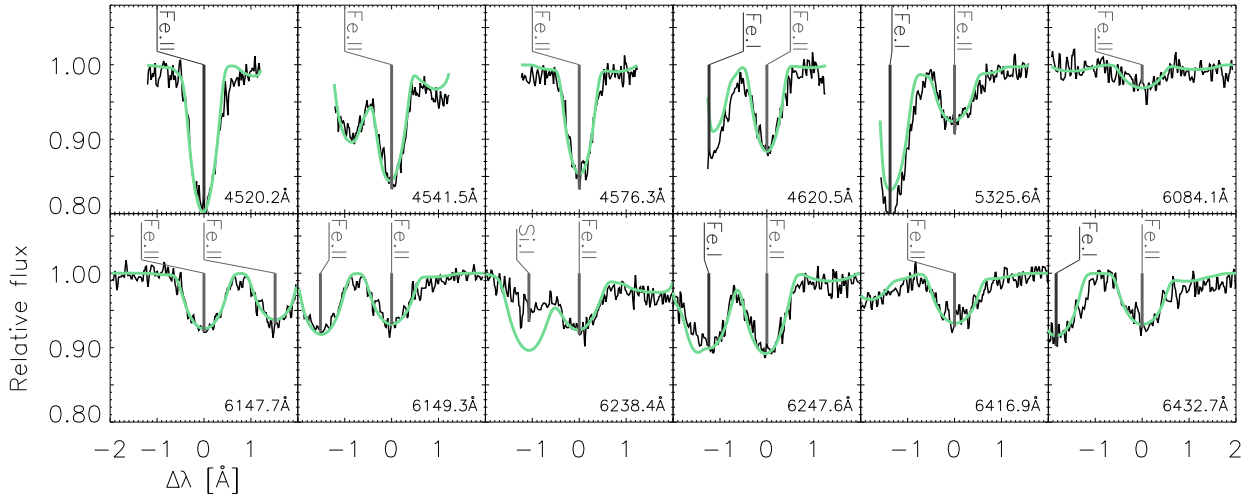}
   \caption{Twelve Fe\scii\ lines in the candidate \gamdor\ star 
HD~110379 fitted by \vwa\ (continuous line).
The wavelengths of the fitted lines are given in the bottom right corner of each panel.
              \label{fig:fitvwa}}
    \end{figure*}

\section{Aims and methods\label{sec:aims}}

The comparison of observed pulsation frequencies with theoretical
models of \gamdor\ stars will provide important insight into 
main sequence stars with convective cores. It is important to constrain
the model space by putting constraints on \teff, \logg, and metallicity
from observations \citep{moya05}. Except for one star,
our targets are single stars and their fundamental parameters 
must be estimated by indirect methods like the calibration 
of photometric indices and spectroscopic analysis.
In the present analyses we rely on the former 
as input for a detailed abundance analysis,
and we seek to improve the estimates of the fundamental parameters.

\subsection{Realistic uncertainty estimates\label{sec:uncer}}

The fundamental physical parameters of stars are mass, radius, and
luminosity. These relate to the effective temperature, \teff, and
surface gravity, \logg. These two parameters, along with the 
metallicity, describe the properties
of the applied model atmosphere in the abundance analysis. 
We adjust the parameters 
of the model to obtain a consistent result, specifically 
by requiring that the abundances measured for Fe lines of neutral
and ionized species---and lines formed at different depths in the
atmosphere---yield the same result. This is done by adjusting not
only \teff, \logg, but also the microturbulence (\vmic), 
which is a crude parametrization of small-scale motions in the gas.
We point out that in these analyses it is actually
the temperature structure we adjust, but the final result we quote
will be the \teff\ and \logg\ of the best-fitting model.
It is thus not a direct measurement of the true values; 
therefore, it is important to assess how the fitted model parameters 
relate to the parameters of truly fundamental stars, \ie\ stars where the properties
are determined by model-independent means \citep{smalley05}.
The primary reference star is the Sun, but binary stars where accurate masses
and radii have been determined are important for extending
the validity of the method to higher temperatures. 
In our analysis we will test our method on the Sun and the 
visual binary HD~110379 (\cf\ Sect.~\ref{sec:fund}).

Realistic uncertainties on \teff\ and \logg\ are especially
important for our target stars, since some of them will likely 
be the targets for detailed asteroseismic studies. 
Our classical analysis of spectral lines
yields intrinsic uncertainties on \teff\ and \logg\ 
in the range $50$--$100$~K and $0.08$--$0.12$~dex for slowly 
rotating stars. We estimate that at least $100$~K and $0.1$~dex
must be added due to the limitations of 
the model atmosphere alone.
This will lead to uncertainties on the derived abundances 
of the order of $0.08$ dex, which must be added to the 
measured intrinsic scatter.

\subsection{Model atmospheres\label{sec:models}}

We used model atmospheres interpolated in the fine grid published by \cite{heiter02}. 
These models are based on the original ATLAS9 code by \cite{kurucz} 
but use a more advanced convection description \citep{kupkaconv} based on \cite{canuto92}.
Our sample consists of A- and F-type stars that have shallow convection zones.


One of the physical assumptions in the models
is local thermodynamical equilibrium (LTE), 
but deviations from LTE start to become important for the hotter stars.
We have not included the NTLE corrections in
the present analysis but will estimate the importance of the effect here.
According to \cite{rent96}, the correction for neutral iron is
[Fe\ione/H]$_{\rm NLTE}=$\,[Fe\ione/H]$_{\rm LTE}+0.1$ dex 
for stars with solar metallicity and \teff\,$=7300$~K.
When this correction is applied, Fe\itwo\ (unaffected by NLTE) must be increased by adding $+0.2$ to \logg.
When extrapolating from Fig.~5 in \cite{rent96}, 
the NTLE effect becomes negligible for stars cooler than about 6,000~K.

The initial model atmosphere used for the abundance analysis of each star
has \teff\ from the $V-K$ colour and \logg\ from the \hipp\ estimates,
except for HD~110379 where we used \logg\ from the binary orbit \citep{smalley02}.
For the metallicity we used the estimate from the \str\ $m_1$ index and
solar metallicity for the two stars 
that did not have this index (HD~65526 and HD~218225).
We note that the photometric amplitudes in variable targets
are so tiny that they will not affect the applied indices.
We used an initial microturbulence of 1.5\,\kms\ for all stars.
The adopted values for the fundamental parameters used for the initial models
are printed in bold face in Table~\ref{tab:fund}.

\section{Abundance analysis with \vwa\label{sec:vwa}}

The software package \vwa\ \citep{bruntt02, bruntt04} was used to 
measure abundances in the spectra and to constrain \teff\ and \logg\ for the
slowly rotating stars. We have expanded \vwa\ so it now has a graphical user interface (GUI),
which allows the user to investigate the spectra in detail, pick lines manually, 
inspect the quality of fitted lines, etc.

Abundance analysis with \vwa\ relies on the calculation of synthetic spectra. 
We use the \synth\ code by \cite{sme}, which works with
\atlasni\ models and atomic parameters and line-broadening coefficients
from the \vald\ database \citep{vald}.
Compared to classical abundance analyses based on equivalent widths,
our analysis has two important advantages:
\begin{itemize}
\item The calculated spectrum includes contributions from neighbouring lines, 
and \vwa\ can analyse stars with high \vsini. 
\item Problems with normalization of the continuum can be recognized 
when comparing the observations with the calculated spectrum. 
\end{itemize}
These effects gradually become stronger when going to shorter wavelengths.

In our experience when \vsini\ becomes
larger than about 50\,\kms, we cannot simultaneously constrain
microturbulence, \teff, or \logg. This is because increased line blending
and improper normalization of the continuum will introduce relatively
large systematic errors (\citealt{erspamer03}; see also Sect.~\ref{sec:synth}).

Each line is fitted by iteratively changing the abundance to match the 
equivalent width (EW) of the observed and calculated spectrum. 
The EW is computed in a wavelength interval equal to the full-width 
half-maximum (FWHM) of the line. In some cases, \eg\ if the line is 
partially blended in one wing of the line, the range for fitting 
the EW must be changed manually in the GUI. 
On a modern computer (3.2~Ghz Pentium~IV),
it takes about one hour to fit 250 lines for a star with low \vsini. 
The fitted lines are inspected in the GUI, problems with 
the continuum level or asymmetries in the line 
are readily identified, and these lines are discarded.  
This is done automatically by calculating the \chitwo\ of the
fit in the core and the wings of the lines. This is followed by
a manual inspection of the fitted lines.

An example of 12 lines fitted with \vwa\ is shown in
Fig.~\ref{fig:fitvwa} for the star \gamdor\ candidate HD~110379. 
The star has a moderately high projected rotational velocity of \vsini~$\simeq25$\,\kms.
It is seen that a few of the lines are 
affected by blends from strong neighbouring lines.
As an example of the line lists we have used, we
list the atomic parameters of the spectral lines we used 
for HD~110379 in Table~\ref{tab:linelist} in Appendix~\ref{sec:app110}.


%
   \begin{figure}
   \centering
 \includegraphics[width=9.4cm]{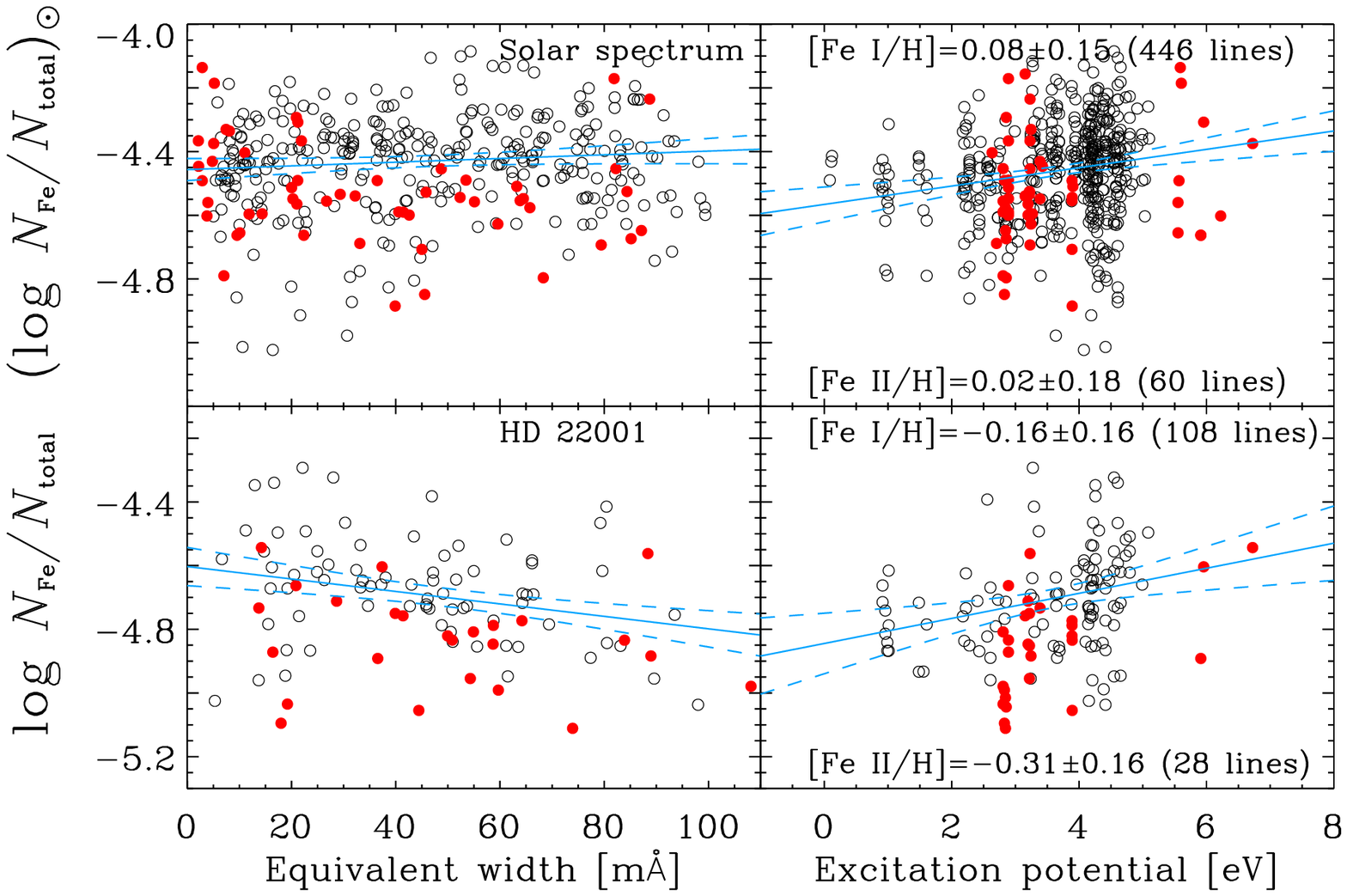}  
 \includegraphics[width=9.4cm]{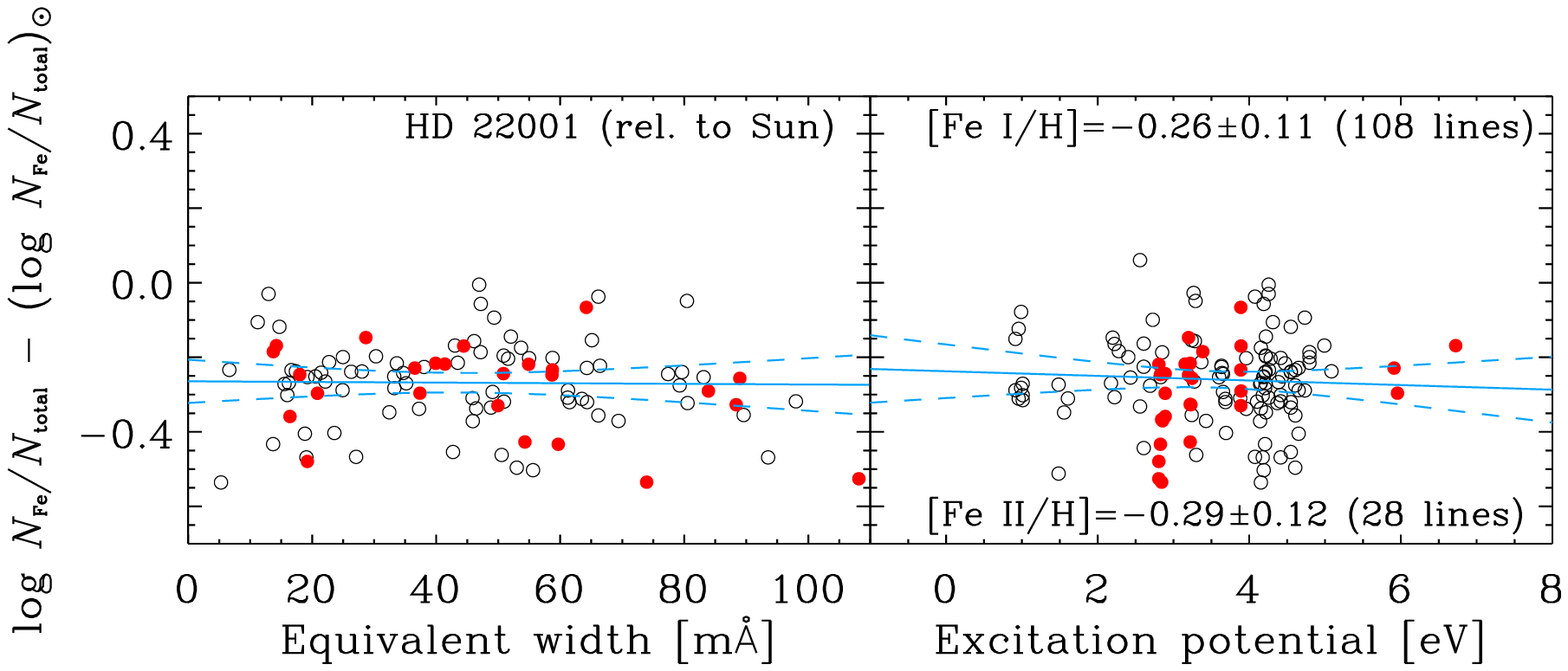}  
\vskip 0.25cm  
   \caption{The two \topp\ \panels\ show the abundances of Fe
in the reference spectrum of the Sun and the target star HD~22001. 
The open symbols are Fe\ione\ lines and solid symbols are Fe\itwo\ lines.
In the \bott\ \panel\ the abundances in HD~22001 are measured relative to the
same lines in the Sun. As a consequence the \rms\ scatter decreases by 40\%.
The solid lines are linear fits and the dashed 
lines are 95\% confidence limits.
              \label{fig:raw}}
    \end{figure}

\subsection{Correcting \logg\ values: relative abundances}

In addition to the sample of stars in Table~\ref{tab:fund}, 
we analysed a reference spectrum of the Sun \citep{hinkle}, which has
high resolution and high signal-to-noise (S/N\,$\simeq1,000$).
Using the results for the solar spectrum allows us to make a more precise 
differential abundance analysis.
The abundances of Fe\ione\ and Fe\itwo\ lines 
measured in the Sun and HD~22001 are compared in the two \topp\ \panels\ in Fig.~\ref{fig:raw}.
The abundances are plotted against equivalent width and excitation potential 
in the \lee\ and \rii\ \panels, respectively.
There are 446 Fe\ione\ lines in the Sun but only 108 lines are available for HD~22001,
mainly because its spectrum has lower S/N
and the star has higher \vsini\ (\ie\ fewer unblended lines), and is about 1,200~K hotter than the Sun.
The \rms\ scatter of the Fe\ione\ abundance is about 0.15 dex for both stars.

It is seen that for the solar spectrum (\topp\ \panels\ in  Fig.~\ref{fig:raw}),
the abundance of Fe from neutral and ionized species do not agree and 
there is a significant positive correlation with excitation potential.
The former could mean that \logg\ is too high, while the latter indicates
that the temperature of the model is too low.
Since \logg\ and \teff\ are well-known for the Sun, 
we can make a first-order correction of the atmosphere models 
by measuring abundances in the target stars relative to the Sun.
When doing this line-by-line any erroneous oscillator strengths, \loggf, are also corrected.
This procedure has been used previously in detailed abundance studies 
of solar-like stars \citep{gonzalez98} and also stars of earlier type \citep{gillon06}.
To give an idea of the magnitude of the \loggf\ corrections, we quote
the \rms\ of the corrections for a few elements: C/Sc/Ni: 0.11 dex, 
O/Ca/Fe lines: 0.18~dex, S/Ti/Cr: 0.23 dex, and for Si: 0.46~dex. 

The result of the differential abundance analysis for HD~22001 is shown 
in the \bott\ \panel\ in Fig.~\ref{fig:raw}.
It is seen that the \rms\ scatter in the Fe\ione\ and Fe\itwo\ lines 
is lower by about 40\%.
While the differential analysis 
improves the internal precision of the measured abundances significantly, 
one should note that our targets are $300$--$1,500$~K hotter than the Sun,
and therefore systematic errors could be the dominant source of uncertainty on the
abundances and the fundamental parameters.
The amount of convection will be quite different
in the sample stars compared to the Sun, and the temperature structure
in these model atmospheres may not describe 
the observed stars correctly \citep{heiter02}.
The fact that the \rms\ scatter decreases significantly gives us
some confidence in the differential analysis, 
but systematic effects on \teff\ or \logg\ could be introduced. 
To explore this caveat, 
we analysed some secondary and tertiary reference stars that
have spectral types similar to our targets.

   \begin{figure}
   \centering
    \includegraphics[width=8.8cm]{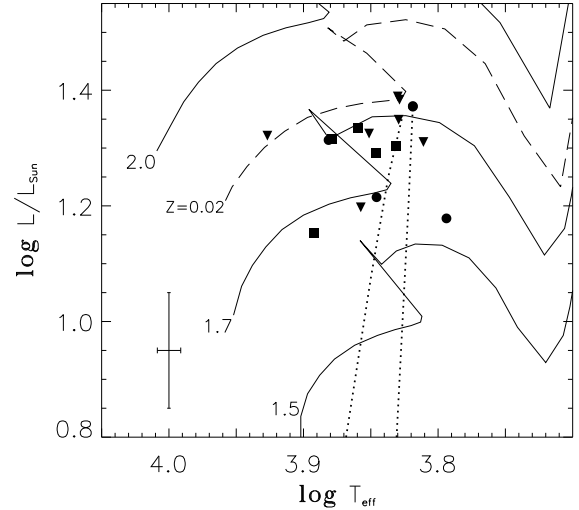}
   \caption{Hertzsprung-Russell diagram for the sample of \gamdor\ candidate stars analysed. 
Circles, boxes, and triangle symbols are used for stars with
low, moderate, and high \vsini.
Three evolution tracks from \cite{lejeune01} are shown for
metallicity $Z=0.008$ (solid lines) and one track with $M/M_\odot=2.0$ for Z=0.02 (dashed line). 
The dotted lines mark the \gamdor\ instability strip predicted from theoretical models by \cite{dupretinstab}.
              \label{fig:hr}}
    \end{figure}

%
   \begin{figure*}
   \centering
   \includegraphics[angle=90,width=17cm]{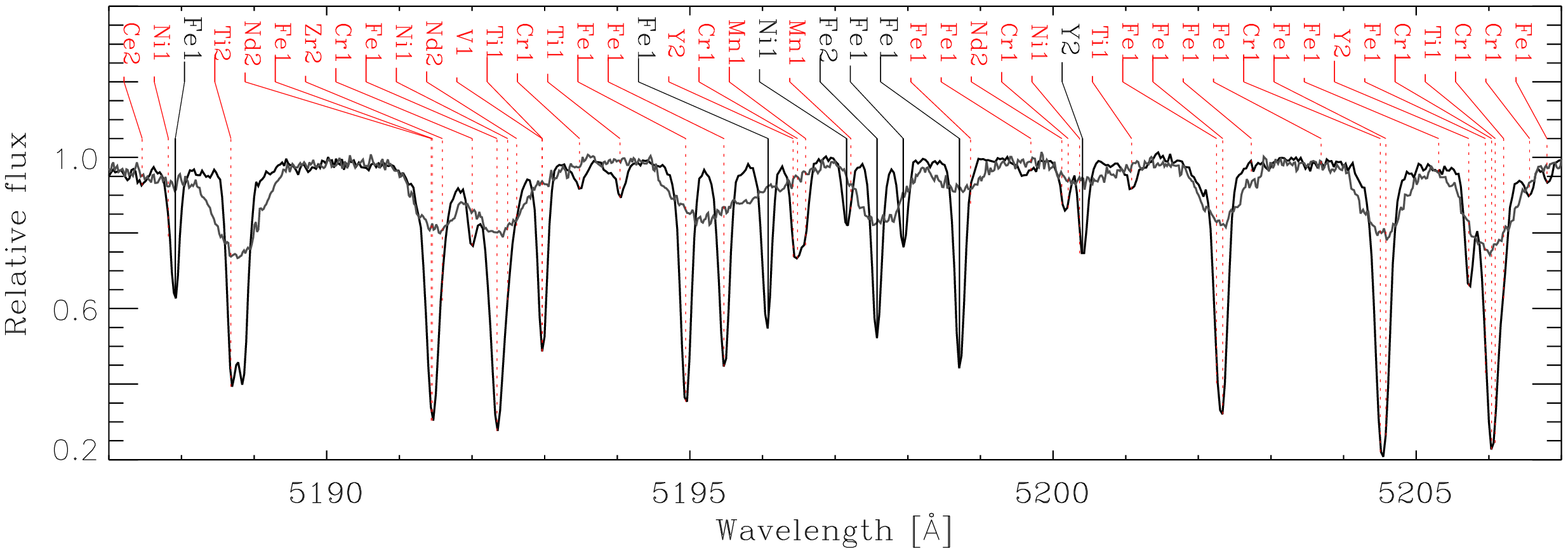}
   \caption{Comparison of the spectra for the Sun and HD~110379 for which
\vsini\ is 2 and~24 \kms.
Atomic lines selected for abundance analysis of the Sun are marked with solid vertical lines,
while the neighbouring spectral lines are marked with dotted lines.
              \label{fig:hxx}}
    \end{figure*}

\subsection{New results for the abundance in the Sun}

Based on time-dependent 3D hydrodynamical models,
updated atomic line parameters, and NLTE corrections,
\cite{grevesse07} recently revised the abundances in the Sun.
The overall metallicity, $Z$, has decreased significantly
from previous estimates \citep{grevesse98}, 
\ie\ from $0.017$ to $0.0122$ ($-30$\%), mainly due to the new C, N, and O abundances.
This result has vast implications in many fields of astrophysics.
This includes detailed asteroseismology of \gamdor\ stars,
which the current work will provide important input to
in terms of fundamental parameters and abundances.
Our analysis was initiated before the new results, so they rely on the
previous solar abundances from \cite{grevesse98}.
We are convinced that the analysis is still valid
since it is carried out differentially with respect to an observed spectrum of the Sun.
We have also did the analysis applying the new solar abundance 
for one of our reference stars, HD~110379, as an explicit check. 
The mean abundance
is $0.02$ dex higher using either Fe\,\ione\ and Fe\,\itwo\ lines,
which is certainly within the uncertainty on the metallicity. 
The derived values for the microturbulence, \teff, and \logg\ are unchanged.

\section{Parameters from photometry and parallaxes\label{sec:fundamental}}

\subsection{\str\ indices and 2MASS $(V-K)$\label{sec:phot}}

We have derived effective temperature (\teff), surface gravity (\logg), 
and metallicity ([Fe/H]) using both \str\ indices and the $V-K$ colour from the 2MASS
point source catalog \citep{cutri03}.
Furthermore, we used parallaxes from \hipp\ to determine \logg.
The fundamental atmospheric parameters of the 
sample of \gamdor\ stars we analysed are given in Table~\ref{tab:fund}. 

We used the \templogg\ software \citep{rogers95} to derive the
fundamental atmospheric parameters from the \str\ indices. 
The on-line version of \templogg\ provides uncertainty estimates that are too optimistic,
since they are solely based on the uncertainties of the photometric indices. 
In Table~\ref{tab:fund} we quote uncertainties on \teff, \logg, 
and \feh\ of 250~K, 0.2 dex, and 0.1 dex \citep{rogers95,kupka01}.

We used Str\" omgren colour indices from the compilation of \cite{hauck98},
but they were not available for two stars: HD~65526 and HD~218225. 
This $H_\beta$ index was not available for HD~12901 and HD~14940
so interstellar reddening, $E(b-y)$, could not be determined.
However, in all cases, $E(b-y)<0.01$ with the exception of 
HD~125081, which has $E(b-y)=0.045\pm0.005$. 
This star also has a high $m_1$ index,
indicating the star is quite metal rich. 
From \templogg, we get \feh\,$=+0.54\pm0.10$, 
but this is based on an extrapolation from the calibration by \cite{olsen88}.
The \str\ indices for HD~110379 listed 
in \simbad\ are incorrect, as also noted by \cite{scardia07}.
Instead we used the average of the indices listed in 
\cite{crawford66}, \cite{cameron66}, and \cite{olsen83}, which are all in good agreement.
The \str\ indices in \cite{hauck98} for HD~7455 are the mean of 
\cite{stetson91} and \cite{perry91}, which are not in agreement:
the difference in the $b-y$ index is $0.086$. 
Using the indices from \cite{stetson91} yields \teff\,$=5800$\,K and \logg\,$=4.9$,
while \cite{perry91} gives \teff\,$=6460$\,K and \logg\,$=4.0$. 




The $V-K$ colours from the 2MASS catalogue were used to estimate \teff\ with
the calibrations from \cite{masana06}. We adopted the interstellar reddening from
\templogg\ and assumed $E(b-y)=0$ for the five stars where it was not available.
We used $E(V-K)=3.8\,E(b-y)$ using \cite{cardelli89}.
The $V-K$ calibration only has a weak dependence
on [Fe/H] and \logg, so we assumed \feh\,$=-0.2\pm0.2$ and \logg\,$=4.0\pm0.3$ for all stars.
This is a valid assumption since the maximum change in \teff\ is 40~K when changing either \feh\ or \logg\ by 2\,$\sigma$. 
The four brightest stars, HD~22001, HD~27290, HD~33262, and HD~110379, have $V<5$, and their 2MASS $K$ 
band magnitudes are based on saturated images. For this reason the errors are large, \ie\ 
$\sigma_K=0.23$\,mag, instead of $\simeq0.02$\,mag for the other stars.
For these stars the uncertainty on \teff\ from the $V-K$ calibration is around $500$~K, while for
the other stars it is around $80$--$100$~K.
We find that \teff\ from the \str\ indices
and $V-K$ agree within the uncertainties except for HD~110379. 
This star is part of the visual binary system $\gamma$~Vir, and the companion, 
which has equal brightness, is within 4 arc seconds \citep{scardia07}.

\subsection{\logg\ from \hipp\ parallaxes\label{sec:hipp}}

The Hertzsprung-Russell diagram for the stars is shown in Fig.~\ref{fig:hr}.
It is based on the adopted \teff\ and luminosities calculated from
the \hipp\ parallaxes and a solar bolometric magnitude of $M_{\rm bol, \odot}=4.75$. 
In Fig.~\ref{fig:hr} we also show the \gamdor\ instability strip from \cite{dupretinstab} based on models 
with a mixing-length parameter $\alpha=2.0$ along with
evolution tracks from \cite{lejeune01} for metallicities $Z=0.008$ and $Z=0.02$, 
which bracket the range for our targets.
Each track is marked by the mass in solar units. From these tracks we
can estimate the masses of the stars to be in the range $1.6$--$2.0$\,\msun.
Since the target stars lie in the region of the ``hook'' of the 
evolution tracks, the uncertainty on the masses is $\simeq0.2$~\msun.
We have assumed a common mass of $1.8\pm0.2$\,\msun\ except for the well-studied
binary star HD~110379, which has a known mass $M/M_\odot=1.4$ \citep{scardia07}. 

We used this mass estimate, the adopted \teff, 
and the \hipp\ parallaxes to determine \logg\ values. 
In particular we used\footnote{This expression is derived from the basic formulae 
$g\propto M/R^2$, $L\propto R^2\,T_{\rm eff}^4$, and the definition of absolute magnitude,
while assuming the solar values $M_{\rm bol}=4.74$ and $\log g_\odot=4.44$.}
$\log g = 4 [T_{\rm eff}] + [M] + 2 \log \pi + 0.4 (V + BC_V + 0.26) + 4.44$,
where $[T_{\rm eff}] = \log (T_{\rm eff}/T_{{\rm eff}\,\odot})$ and 
$[M] = \log (M/M_\odot)$.
We used bolometric corrections ($BC_V$) from the tables by \cite{bessell98}.
If we assume the mass is known to 10\% and \teff\ to 4\% for all stars in the sample, 
the uncertainty on \logg\ will depend on the uncertainty of
the parallax: 13 out of 18 stars have uncertainties below 7\%, while
five stars, HD~7455, HD~26298, HD~125081, HD~126516, and HD~218225, have
uncertainties around 15\%. The uncertainty on \logg\ is 0.13\,dex 
and 0.20\,dex for these two groups of stars.
The \logg\ values from the Str\" omgren $c_1$ index          
and the \hipp\ parallaxes 
agree, but the estimated uncertainty on the latter 
is significantly lower: typical uncertainties on
\logg\ are $0.2$ and $0.1$~dex, respectively.



\begin{table}
\caption[]{Fundamental parameters obtained with \vwa\ for the reference stars 
after perturbing the initial value of either \teff\ or \logg.
\label{tab:pert}}
\centering                          
\begin{tabular}{rr|lll}
\hline\hline
Star     & Pert.\   & \teff & \logg & [Fe/H] \\
\hline	    

Sun       & \ittt & $ 5760\pm25  $  & $4.46\pm0.03$ & $  -0.03\pm0.05$  \\
          & \iggg & $ 5780       $  & $4.51       $ & $  -0.01       $  \\ \hline
HD32115   & \ittt & $ 7630\pm140 $  & $4.41\pm0.09$ & $  +0.10\pm0.11$  \\
          & \iggg & $ 7710       $  & $4.47       $ & $  +0.13       $  \\ \hline
HD37594   & \ittt & $ 7440\pm160 $  & $4.18\pm0.12$ & $  -0.26\pm0.13$  \\
          & \iggg & $ 7310       $  & $3.97       $ & $  -0.33       $  \\ \hline
HD49933   & \ittt & $ 6770\pm90  $  & $4.24\pm0.08$ & $  -0.43\pm0.09$  \\
          & \iggg & $ 6780       $  & $4.24       $ & $  -0.42       $  \\ \hline
HD110379  & \ittt & $ 7070\pm130 $  & $4.03\pm0.09$ & $  -0.04\pm0.15$  \\
          & \iggg & $ 7170       $  & $4.27       $ & $  +0.00       $  \\ \hline

\multicolumn{5}{l}{Abbreviations used in second column:}\\
\multicolumn{5}{l}{\ittt $=$ perturbation of \teff\,$=\pm300$\,K}\\
\multicolumn{5}{l}{\iggg $=$ perturbation of \logg\,$=\pm0.4$\,dex}\\

\end{tabular}
\end{table}

\section{Fundamental parameters from spectroscopy\label{sec:fund}}

Two important aims of the current work are to determine 
the fundamental parameters of the \gamdor\ stars and 
to compare their abundance pattern with other stars
of similar spectral type. 
We first analyse a few stars with well-known parameters. 
We then analyse synthetic
spectra in order to estimate uncertainties and make sure our method
can be used to reliably constrain \teff, \logg, and metallicity.
Based on these results we will decide on the
approach for the detailed analysis of the target stars.


\vwa\ can automatically adjust the microturbulence, \teff, and \logg\ either
simultaneously or any parameter can be fixed. 
This part of the analysis is based only on Fe lines,
which are the most numerous in the stellar spectra.
The iterative process of adjusting the parameters is to

\begin{itemize}
\item Minimize the correlation of Fe\ione\ abundance with equivalent
width. This is done by adjusting the microturbulence.
\item Minimize correlation between Fe\ione\ abundance and lower excitation
potential of the atomic level causing the line. This is done by
adjusting \teff.
\item Minimize the difference in abundance of Fe\ione\ and Fe\itwo\ lines.
This is done by adjusting \logg.
\end{itemize}

Unfortunately, microturbulence and \teff\ are not independent. 
Therefore, only the most sensitive lines are used to adjust the microturbulence
(typically EW $< 80$\,m\AA), while we also use stronger lines ( EW $< 150$\,m\AA) 
to constrain \teff. 
For the slowly rotating stars, \vwa\ needs to run $5$--$8$  iterations before it converges. 
In each iteration, up to 120 lines are used and the CPU time 
is typically $1$--$3$ hours for each star. 

An important limitation to detailed spectroscopic analyses arises when
the lines are broadened due to rotation.
In Fig.~\ref{fig:hxx} we show part of the spectrum for the CORALIE spectra
of the Sun and the selected of lines for the abundance analysis. The spectrum
of HD~110379, which has \vsini\,$=25$\,\kms, is also shown for comparison.
It is seen that line blending is worse, which illustrates that correct placement 
of the continuum can be difficult as \vsini\ increases. 
We will assess the importance of rotation by analysing 
synthetic spectra with increasing \vsini\ below.

\begin{table}
\caption[]{Uncertainties on the fundamental parameters from the \vwa\ analysis of synthetic spectra with 
two values of \teff\ and with increasing \vsini\ for one value of \teff.
\label{tab:synth}}
\centering                          
\begin{tabular}{cc|ccc}
\hline\hline
\multicolumn{2}{c|}{Model parameters}  & \multicolumn{3}{c}{Uncertainties} \\
\vsini\ [\kms]  & \teff\ [K]  & $\sigma($\teff$)$ [K] & $\sigma($\logg$)$ & $\sigma($\feh$)$ \\ \hline
10              & 7250        & 40                    & 0.03              & 0.02              \\
10              & 6750        & 50                    & 0.04              & 0.02              \\
20              & 6750        & 110                   & 0.08              & 0.03              \\
40              & 6750        & 140                   & 0.11              & 0.03              \\
60              & 6750        & 200                   & 0.13              & 0.04              \\ \hline

\hline

\end{tabular}
\end{table}

\subsection{Primary reference: solar spectrum from CORALIE\label{sec:prim}}

We used \vwa\ to analyse a spectrum of the Sun measured with CORALIE. 
The S/N was 180, which is slightly higher than the spectra for the target stars.
We ran the software with four models with parameters offset 
by $\pm300$~K in \teff\ and $\pm0.4$~dex in \logg, respectively. 
In Table~\ref{tab:pert} we compare the results for the derived fundamental parameters. 
The results are very close to the canonical values 
of \teff\,$=5777$~K, \logg\,$=4.44$, and \feh\,$=0.00$.
The quoted uncertainties on \teff\ and \logg\ were estimated as is
described in Sect.~\ref{sec:finalfund} and the quoted uncertainty on \feh\ is
the \rms\ value of the abundance determined from the Fe\,\ione\ lines.
Uncertainties on \teff\ are rounded off to 10~K. 
The uncertainties in Table~\ref{tab:pert} do not include any contribution from the uncertainty on the atmosphere models (\cf\ Sect.~\ref{sec:uncer}).

\subsection{Secondary reference: the visual binary star HD~110379\label{sec:seco}}

HD~110379 is the A component in the visual binary system $\gamma$~Vir, 
which has two identical components \citep{popper80}.
From the orbital mass, the \hipp\ parallax, and measured spectrophotometric fluxes, 
constraints can be placed on \logg\ and \teff. 
Thus, \cite{smalley02} include HD~110379 in their sample of fundamental stars and derived 
\teff\,$=7143\pm450$\,K and \logg\,$=4.21\pm0.02$. 

We used HD~110379 to test the robustness of results from \vwa.
The S/N in the spectrum is 120, and the star has 
\vsini\,$=24$\,\kms\ (part of the spectrum is shown in Fig.~\ref{fig:hxx}).
For the initial model we used the \teff\ and \logg\ estimate from the \str\ indices and the \hipp\ parallax, 
\ie\ \teff$\,=6860$\,K and \logg$\,=4.39$ (\cf\ Table~\ref{tab:fund}). 
We also perturbed the initial guess for the fundamental parameters
and converged at the parameters listed in Table~\ref{tab:pert}.
One of the results (p\logg) is in excellent agreement with \logg\ from the binary orbit,
while the other result (p\teff) has lower values of both \teff\ and \logg.
This could be an indication that the determined values of \logg\ and \teff\ are not independent,
although the results for the tertiary reference stars, 
except perhaps for HD~37594, do not indicate that this is a general problem.

\subsection{Tertiary references: HD~32115, HD~37594, HD~49933\label{sec:tert}}

We analysed CORALIE spectra of three slowly rotating 
A- and F-type stars for which detailed analyses have been published:
HD~32115, HD~37594 \citep{bikmaev02}, and HD~49933 \cite{gillon06}.
The fundamental parameters in these stars are constrained
by photometric indices and spectroscopic analysis. 
Therefore, 
\logg\ is known to about $0.15$~dex from the parallax,
which is about an order of magnitude worse than for the primary and secondary reference. 
The S/N in the three spectra from CORALIE is 200, 220, and 140 and
the stars have \vsini\,$=9$, $17$, and $10$\,\kms.

As for the primary and secondary reference stars, 
we offset the initial parameters to test the convergence of \vwa. 
The results are shown in Table~\ref{tab:pert}. 
It is encouraging that for each star,
the results agree within the error bars.
The two slowly rotating A-type stars HD~32115 and HD~37594 were analysed by \cite{bikmaev02},
who adopted a {\em fixed} value for \teff\ based on \str\ indices and the H$\alpha$ line and 
\logg\ from the \hipp\ parallax.
The F-type star HD~49933 was analysed by \cite{gillon06}, who used
an approach similar to \vwa\ to fit \teff\ and \logg\ as part of the analysis.

Our results are in acceptable agreement with previous studies.
The metallicities agree within 0.1 dex, while the largest difference 
in \teff\ and \logg\ are $200$~K and $0.2$. 
The differences are largest for HD~32115 and HD~37594, but we recall that
\teff\ and \logg\ were not adjusted as part of the abundance analysis by \cite{bikmaev02}.
On the other hand, the agreement is good for HD~49933, in which 
case the \vwa\ analysis is quite similar to the approach of \cite{gillon06}. 
We recall that we did not include NTLE effects, although for the
two hottest stars, the effect on \logg\ would be about $+0.2$~dex. 
However, the studies we are comparing with here also did not include 
any correction.

   \begin{table}
      \caption[]{The three groups of stars, depending on their \vsini.
         \label{tab:groups}}
\centering                          
\begin{tabular}{l|l|l|l}
\hline\hline
        & Low \vsini         & Moderate        & High \\ 
\hline
\bon    & 167858             &  14940,  40745, &  12901, 27290,      \\
        &                    &  48501, 135825  &  65526,218225       \\ \hline
\can    & 110379, 126516     &                 &  26298              \\ \hline
\bfs    & 125081             &                 &                     \\ \hline
\cons   & 7455, 22001,       &                 & 27604, 85964        \\         
        & 33262              &                 &                     \\  \hline 
\reff   & Sun (CORALIE),     &                 &                     \\
        & 32115, 37594,      &                 &                     \\ 
        & 49933              &                 &                     \\ \hline 
Free param.  & \vmic, \teff, \logg & \vmic, \teff   & \teff  \\ \hline
\multicolumn{4}{l}{The first column is the variability type from \cite{decat06}.}\\
\multicolumn{4}{l}{The free parameters in the analysis are given below each group.}\\ 

\end{tabular}
\end{table}

\subsection{Abundance analysis of synthetic spectra\label{sec:synth}}

We tested \vwa's ability to determine \teff, \logg, and metallicity 
by using synthetic spectra with the \synth\ code. This is the ``ideal'' case
for abundance analysis since all \loggf\ values are known and the spectrum is correctly
normalized by design.
Also, the input fundamental parameters are known.
To mimic the quality of the observed data, we added random noise 
corresponding to S/N\,$=100$ in the continuum. 
We calculated spectra with \teff\,$=6750$ and $7250$~K and \logg$\,=4.3$.
For the cooler model, we used a range in \vsini\ of 10--60 \kms, and for the
hotter one, we used $v \sin i=10$\,\kms. The spectra were calculated in the
range $4500$--$5600$\,\AA\ where most of the lines are present. 
For slow rotation ($v \sin i <20$), we used about 100 and 20 lines 
of Fe\ione\ and Fe\itwo\ in the analysis.
For the fast rotation (40 and 60 \kms), only half as many lines were used.

We offset the initial models in \teff\ ($\pm500$~K) or \logg\ ($+0.4$~dex) and
let \vwa\ determine the best parameters.
For slow and moderate rotation, $v \sin i =10$--$40$\,\kms, 
we found that the models converged satisfactorily:
the largest difference in \teff, \logg, and \feh\ were $30$~K, $0.05$~dex, and $0.03$~dex.
For stars with high $v \sin i =60$\,\kms, we found 
$\Delta T_{\rm eff}=60$~K, $\Delta \log g=0.05$, $\Delta {\rm [Fe/H]}=0.08$.

We also calculated the uncertainties on \teff\ and \logg\ from 
the analysis of the synthetic spectra, and we list the results in Table~\ref{tab:synth}.
Uncertainties on the fundamental parameters from the models are not included (\cf\ Sect.~\ref{sec:uncer}).
The last column gives the \rms\ value of the Fe\ione\ and Fe\itwo\ abundance. 
In comparison, the uncertainties for the observed secondary and tertiary reference stars 
listed in Table~\ref{tab:pert} are roughly twice as large. 
The reasons are likely a combination of imperfect continuum normalization, 
the remaining errors in the oscillator strengths, and differences in
the temperature structure in the models and the real stars.
The reference stars are all slowly rotating stars, but 
we may, as a first approximation, scale the uncertainties for the ideal case in
Table~\ref{tab:synth} by a factor two. Thus, for stars with $v \sin i>40$\,\kms,
the uncertainties become larger than the estimates from photometric indices
or the \hipp\ parallax. We have therefore chosen not to use \teff\ and \logg\
as free parameters for stars with $v \sin i >40$\,\kms. We use this
result when defining our strategy for the analysis of the target stars.


%
   \begin{figure}
   \centering
\hskip -0.5cm
\includegraphics[width=9.5cm]{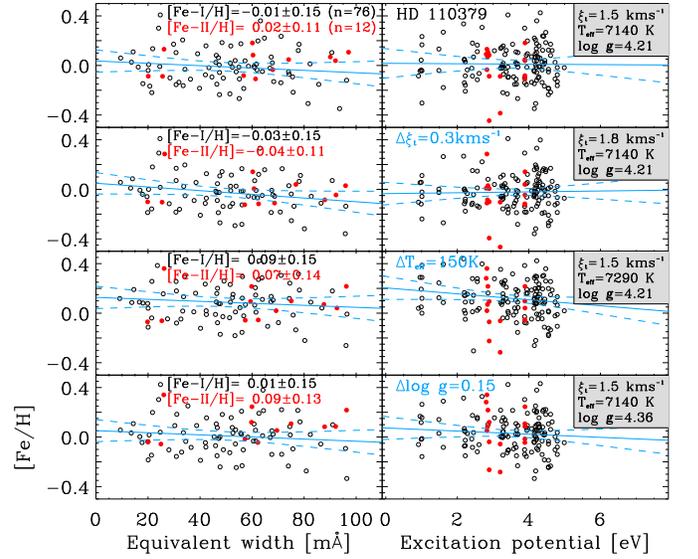}
   \caption{Abundance of Fe in HD~110379 for four different atmosphere models. 
The \topp\ \panel\ is for the adopted parameters, while the next three \panels\ are
for increased \vmic, \teff, and \logg\ (as indicated in the \rii\ \panels).
Fe\ione\ and Fe\itwo\ lines are plotted with open and solid points, respectively.
The solid line is a linear fit to the Fe\ione\ lines and the dashed lines indicate
the 95\% confidence limit of the fit.
              \label{fig:pert}}
    \end{figure}

%
\subsection{Not all stars are equal: Low, moderate, and high \vsini\label{sec:group}}

Based on the analysis of the reference stars and the synthetic spectra,
we have put the stars in three groups depending on their \vsini\ value
in Table~\ref{tab:groups}:
low (4--25 \kms), 
moderate (35--40 \kms), and 
high \vsini\ (50--70 \kms).
Furthermore, the stars are sorted according to the variability 
type from \paper, \ie\ either constant, \candd, \bonn, or \bfss\ stars.

The procedure for analysis with \vwa\ depends on which group the star belongs to:

\begin{itemize}
\item For stars with low \vsini, we allowed microturbulence, 
\logg\ and \teff\ as free parameters (seven stars).
\item For stars with moderate rotation, we fixed \logg\ from the \hipp\ parallaxes 
(the uncertainty is below 0.15 dex),
but adjusted \teff\ and microturbulence of the atmospheric models (four stars). 
\item For stars with a high rotation rate, there are not enough unblended lines to be able 
to constrain the microturbulence, \logg, and \teff. 
We fixed the value of the microturbulence ($=1.5$ \kms),
used \logg\ from the calibration using the \hipp\ parallax, and fitted only \teff\ (seven stars).
\end{itemize}

\begin{table}
\caption[]{Fundamental parameters determined with \vwa\
for the 18 target stars and the four reference stars.
\label{tab:vwafund}}
\setlength{\tabcolsep}{5pt}
\centering                          
\begin{tabular}{r|crcr} 
\hline\hline
       &        &                            &       & \multicolumn{1}{c}{\vmic\       } \\ 
     HD& \teff  & \multicolumn{1}{c}{\logg}  & \meh  & \multicolumn{1}{c}{       [\kms]} \\ 
\hline	    

  7455 &          $5840\pm120$ &         $4.63\pm0.14$& $-0.38\pm0.08$ &     $1.0\pm0.3$ \\ 
 12901 &          $6950\pm220$ & $^{\rm F}4.07\pm0.13$& $-0.13\pm0.16$ & $\fxx1.5\pm0.4$ \\ 
 14940 &      $    7380\pm180$ & $^{\rm F}4.25\pm0.12$& $+0.01\pm0.09$ &     $1.8\pm0.3$ \\ 
 22001 &          $7010\pm160$ &         $4.19\pm0.14$& $-0.22\pm0.08$ &     $2.6\pm0.3$ \\ 
 26298 &       $   6790\pm200$ &         $3.95\pm0.22$& $-0.27\pm0.11$ & $\fxx1.5\pm0.5$ \\ 
 27290 &       $   7120\pm200$ & $^{\rm F}4.29\pm0.18$& $+0.05\pm0.14$ & $\fxx1.5\pm0.5$ \\ 
 27604 &       $   6320\pm220$ &         $3.65\pm0.24$& $+0.14\pm0.08$ & $\fxx1.5\pm0.5$ \\ 
 33262 &          $6440\pm150$ &         $4.69\pm0.16$& $-0.08\pm0.09$ &     $1.3\pm0.4$ \\ 
 40745 &          $6840\pm180$ & $^{\rm F}4.05\pm0.12$& $-0.00\pm0.09$ &     $1.7\pm0.4$ \\ 
 48501 &          $7240\pm190$ & $^{\rm F}4.28\pm0.12$& $+0.15\pm0.11$ &     $1.5\pm0.4$ \\ 
 65526 &        $  7170\pm210$ & $^{\rm F}4.40\pm0.13$& $-0.26\pm0.13$ & $\fxx1.5\pm0.5$ \\ 
 85964 &        $  6600\pm220$ & $^{\rm F}4.14\pm0.13$& $+0.11\pm0.11$ & $\fxx1.5\pm0.5$ \\ 
110379 &          $7140\pm160$ & $^{\rm F}4.21\pm0.02$& $-0.06\pm0.09$ & $1.5\pm0.4$ \\ 
125081 &          $6670\pm140$ &         $3.02\pm0.17$& $-0.29\pm0.09$ &     $2.8\pm0.4$ \\ 
126516 &          $6590\pm120$ &         $4.01\pm0.15$& $-0.19\pm0.08$ &     $1.9\pm0.3$ \\ 
135825 &          $7050\pm180$ & $^{\rm F}4.39\pm0.13$& $+0.13\pm0.09$ &     $1.5\pm0.4$ \\ 
167858 &          $7610\pm150$ &         $4.35\pm0.19$& $+0.22\pm0.08$ &     $1.8\pm0.3$ \\ 
218225 &        $  6920\pm220$ & $^{\rm F}4.31\pm0.21$& $+0.57\pm0.20$ & $\fxx1.5\pm0.5$ \\ \hline 


   Sun & $5770\pm100$ &         $4.49\pm0.10$& $-0.04\pm0.08$ & $0.6\pm0.1$ \\  
 32115 & $7670\pm170$ &         $4.44\pm0.13$& $+0.08\pm0.08$ & $2.4\pm0.2$ \\  
 37594 & $7380\pm190$ &         $4.08\pm0.16$& $-0.31\pm0.08$ & $2.5\pm0.3$ \\  
 49933 & $6780\pm130$ &         $4.24\pm0.13$& $-0.46\pm0.08$ & $1.8\pm0.2$ \\ \hline 

\multicolumn{5}{l}{Parameters marked by an $F$ were held fixed in the analysis. Each }\\
\multicolumn{5}{l}{uncertainty includes contributions from the model as described in}\\
\multicolumn{5}{l}{ the text. The \logg\ value for HD~110379 is from \cite{smalley02}.}

\end{tabular}
\end{table}


\section{Fundamental parameters\label{sec:finalfund}}

In Table~\ref{tab:vwafund} we list the final fundamental 
parameters of the 18 targets stars and the four reference stars. 
We list the average values from Table~\ref{tab:pert} for the reference stars.
Note that for the moderate and
fast rotators some of the parameters were held fixed 
and these are marked by F in Table~\ref{tab:vwafund}, 
\eg\ \logg\ from the \hipp\ parallax.

Uncertainties on the fundamental parameters were estimated 
by evaluating the sensitivity to the changes in microturbulence, \teff, and \logg.
In Fig.~\ref{fig:pert} we show examples for HD~110379. The \topp\ \panel\ is for
the final parameters and the following panels are for 
increased \vmic, \logg, and \teff, respectively. 
Following \cite{gillon06}, the uncertainty on \teff\ is found by multiplying the
change in \teff\ by the ratio of the uncertainty of the slope and the change in slope, $s$, \ie\
$\sigma(T_{\rm eff}) = \Delta T_{\rm eff} \cdot {{\sigma(s)}/{\Delta(s)}}$.
This uncertainty is added quadratically to the estimated 
uncertainty from the model atmospheres as was discussed in Sect.~\ref{sec:uncer}.

The metallicity, \meh, is computed as the average of the 
five metals Ca, Sc, Ti, Cr and Fe for both neutral and ionized lines, 
with the requirement that at least five lines
were used for any element. 

\begin{figure}
\hskip -0.3cm
   \includegraphics[width=9.6cm]{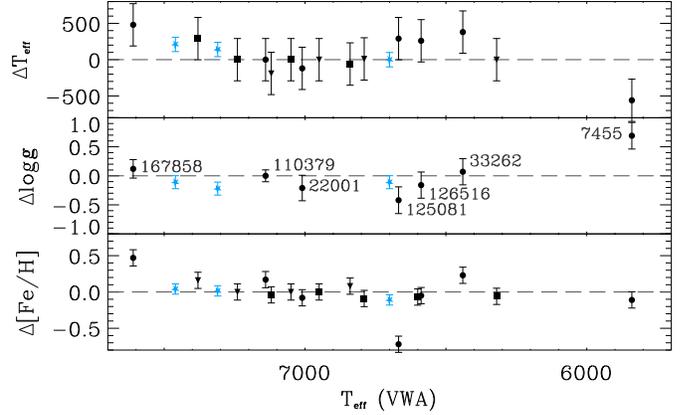}
   \caption{Comparison of the fundamental parameters found
from \vwa\ and photometric indices. Circles, squares, and triangles
are used for the slow, moderate, and fast rotators.
Star symbols are used for the three tertiary reference stars.
The HD numbers of the slow rotators are indicated in the \midd\ \panel.
              \label{fig:vwafund}}
    \end{figure}

In Fig.~\ref{fig:vwafund} we show the differences between
the parameters from \vwa\ and the initial parameters (\cf\ Table~\ref{tab:fund}). 
It is seen that some of the moderate and fast rotators 
have $\Delta T_{\rm eff}=0$~K because they are not very sensitive to changes in \teff.
For the slowly rotating stars, we find that in most cases
\teff\ and \logg\ found by \vwa\ is close to the initial model.
A few exceptions are found that illustrate the importance of
using more than one method to estimate the fundamental parameters.

The largest deviation
is for HD~7455 where \teff\ was 600~K lower and \logg\ 0.7 dex higher 
than the initial model. Our result resolves the dispute
over the discrepant \str\ indices from the two different sources
mentioned in Sect.~\ref{sec:phot}.
We find a large discrepancy for HD~125081, where we get a \logg\ and metallicity 
that is $0.4$~dex and $0.7$ lower, respectively.
This is the most evolved star and is also the only star with a significant 
interstellar reddening. If there was no interstellar reddening, \feh\ would
be lower but not as low as we find from the abundance analysis.
We find a high \teff\ and high metallicity for HD~167858,
but the uncertainty on \teff\ is quite large.

%
%
     \begin{figure*}
     \centering
      \includegraphics[width=4cm,angle=90]{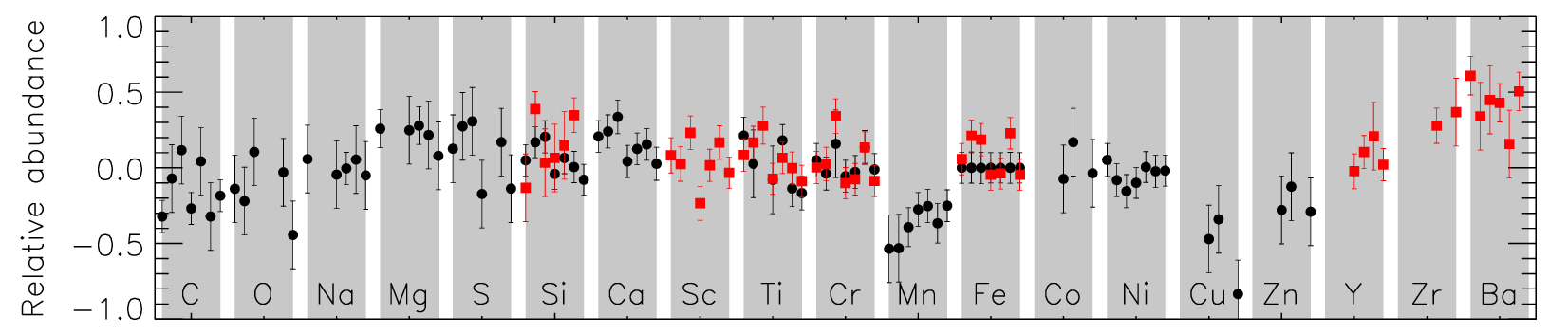}
      \includegraphics[width=4cm,angle=90]{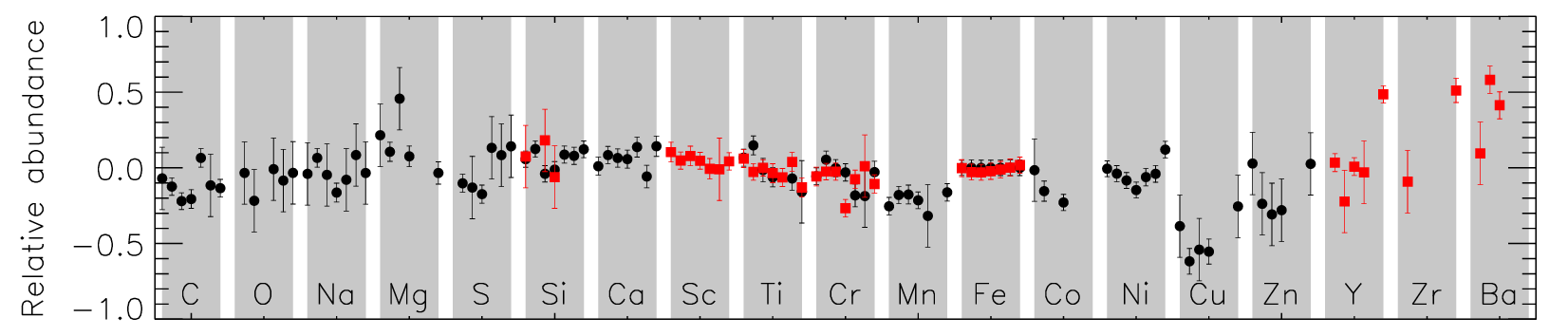}
     \caption{Abundance pattern in the bona fide and three candidate $\gamma$~Dor stars (\topp\ \panel)
              and reference stars and constant stars (\bott\ \panel). 
              Only the results for slow and moderate rotators are shown. 
              The abundances are plotted in the same order from left to right 
              as the HD numbers in Tables~\ref{tab:ab} and \ref{tab:ab2}.
              Note that the abundances are measured 
              relative to the abundance of Fe\ione\ measured in each star.
                \label{fig:abund}}
      \end{figure*}

\section{Abundances}

The abundances for the \bonn\ and \candd\ stars are shown in the
\topp\ \panel\ in Fig.~\ref{fig:abund}, and results for the \cons\ and
\reff\ stars are shown in the \bott\ \panel. 
Results are only shown for the slow and moderate rotators.
For each element, each point corresponds to the HD numbers 
in the same order from left to right in Tables~\ref{tab:ab} and \ref{tab:ab2}.
Note that the abundance of each element has been 
offset by the abundance of Fe\ione,
which is our primary metallicity indicator. 
When this offset is applied we see the abundance pattern quite
clearly, especially for the reference stars where nearly all 
abundances lie within $0.0\pm0.2$~dex.    
In Appendix~\ref{sec:appab} 
we list the individual abundances for all the stars we have analysed.

\subsection{Abundance pattern in the constant stars}

The \bott\ \panel\ in Fig.~\ref{fig:abund}
shows the results for the reference stars and the constant stars in the sample.
We see systematic offsets of about
$-0.5$ to $+0.2$ for C, Mn, Cu, Zn, and Ba, which
to some extent may be explained by the assumption of LTE. 
However, for C the LTE correction for stars
around 7000~K is {\em negative} and of the order of $-0.1$~dex \citep{rent96}.
We included only the line transitions available in \vald,
while for certain elements, like Ba \citep{mcwilliam98}, hyperfine structure is important.

For the Fe-peak elements the scatter is quite low, 
while the scatter from star to star is higher for the lighter elements, C, O, Na, Mg, and S. 
For the light elements, C to S, typically $1$--$5$ lines are available
for each element (\cf\ Table~\ref{tab:ab2}) and so the
uncertainties are quite large due to any systematic errors 
due to erroneous continuum placement, blends, etc.

\subsection{Abundance pattern in the $\gamma$~Doradus stars} 

The \topp\ \panel\ in Fig.~\ref{fig:abund} shows the abundance
pattern for the \bonn\ and \candd\ stars.
We see systematic offsets for C, Mn, Cu, Zr, and Ba, which is
also seen for some elements in the reference and constant stars.
For a given element we see that the scatter is larger than
for the reference stars. However, four of the eight 
stars have moderate \vsini, so fewer lines are available.

One of our goals of the present study is to find evidence of
a link between chemical peculiarity and the \gamdor\ stars.
In particular we searched for evidence of the following patterns:
\begin{itemize}
\item \lamboo: low abundance of Fe-peak elements, solar abundance
of C, N, O, and S.
\item Am: high abundance of Fe-peak elements, low abundance of Ca and Sc.
\end{itemize}
We do not find clear evidence of these patterns in any of the stars.
It was noted by \cite{handler99} that all 
\gamdor\ candidates have metallicities close to the solar value,
based on the \str\ $m_1$ index. 
This is supported by our detailed spectroscopic analyses, since
\meh\ lies in the range $-0.27$ to $+0.22$
for 11 \gamdor\ candidates (see Table~\ref{tab:vwafund}).
The exception is the fast rotator HD~218225, but its value also
has the largest uncertainty, \meh\,$=0.57\pm0.20$.

\subsection{Comparison with previous studies}

Abundance studies have been done previously for three of the stars in our sample, 
HD~48501, HD~110379, and HD~167858.
\cite{boes87} analysed HD~48501 and found \feh\,$=+0.01$ and \cah\,$=+0.20$,
while we get \feh\,$=-0.08$ and \cah\,$+0.23$, which is in very good agreement.
Our results also roughly agree with \cite{boes86}, who found high metallicity in 
HD~167858 at \feh\,$=+0.15$ and \cah\,$=+0.17$, while we find \feh\,$=+0.27$ and \cah\,$+0.32$.
These two studies were based on relatively few lines in a limited optical range.
\cite{erspamer03} analysed several elements in HD~110379, and 
we have good agreement with differences below $0.1$~dex, 
although two elements, Mg and Sc, differ by $0.2$~dex.

Detailed asteroseismic modelling was attempted for the \bonn\ star HD~12901 by \cite{moya05}.
Their analysis was hampered by the uncertain metallicity of about \feh\,$=-0.4$
found from the \str\ $m_1$ index.
We find a metallicity of ${\rm [M/H]}=-0.13\pm0.16$, which we recommend using
in future asteroseismic analyses of HD~12901.
The star is a fast rotator with $v \sin i \simeq 64$\,\kms, so we cannot constrain 
\teff\ and \logg\ based on our analysis with \vwa.

Two stars in our sample are included in the catalogue of Ap and Am stars compiled 
by \cite{renson91}:
HD~125081 is listed as a chemically peculiar star with abnormal abundances of Sr, Cr, and Eu.
HD~167858 is noted as having a ``doubtful nature'', but the source of this claim is not given.
\cite{paunzen98lamboo} did not find any strong chemical peculiarity in these two stars
based on their measurements of the $\Delta p$ photometric index.
Our present analyses of the stars support this result.

%



\section{Conclusions}

We have presented a detailed abundance analysis of a group of
bona fide and candidate \gamdor\ stars. In addition we analysed
a number of constant stars with similar stellar parameters. 
There seems to be larger scatter in the abundances for the \gamdor\ stars,
but we find no strong evidence that the overall abundance pattern 
is different from other A- and F-type stars. 
Furthermore, the metallicity is quite close to the solar value in all cases.
We have constrained the fundamental parameters of 18 single field stars 
from \paper, of which about half are potential \gamdor\ stars. 

We also analysed a few reference stars in order to
thoroughly test the performance of the \vwa\ software package. 
The software gives reliable results for the value of \teff\ and \logg\ 
for our primary and secondary reference stars, 
\ie\ the Sun and the astrometric binary HD~110379;
the latter has a well-determined \logg\ but poorly determined \teff. 
Our results also agree well with previous analyses of three tertiary reference A- and F-type stars,
although these single field stars do not have well-determined values of \teff\ and \logg.
Our analysis of synthetic spectra with increasing rotational velocity 
shows that, for stars with $v \sin i > 50$\,\kms, our method cannot
be used to constrain the microturbulence, \teff, and \logg\ simultaneously.
For the slowly rotating stars with $v \sin i < 25$\,\kms, we can 
constrain \teff, \logg, and \feh\ to about 120~K, 0.13~dex, and 0.09~dex
including estimated uncertainties of the applied model atmospheres.
These results are certainly an improvement 
over photometric uncertainties, which are typically at least twice as large.
We expect that our results will be useful in future asteroseismic studies of \gamdor\ stars.

\begin{acknowledgements}
The project was supported by the Australian and Danish Research Councils and
by the Research Council of Leuven University under grant GOA-2003/04.
This research has made use of the \simbad\ database,
operated at the CDS, Strasbourg, France.
We used atomic data extracted from the \vald\ data base 
made available through the Institute of Astronomy in Vienna, Austria. 
This publication makes use of data products from the Two Micron All Sky Survey, which is a joint project of the University of Massachusetts and the Infrared Processing and Analysis Center/California Institute of Technology, funded by the National Aeronautics and Space Administration and the National Science Foundation.
We are grateful to Friedrich Kupka for his very useful suggestions and
to Fabien Carrier for providing the spectrum of the Sun from CORALIE.
We thank the referee, Patrick Fran{\c c}ois, for his very useful comments.
\end{acknowledgements}

\bibliographystyle{aa}
\bibliography{ms8523} 

\begin{thebibliography}{61}
\expandafter\ifx\csname natexlab\endcsname\relax\def\natexlab#1{#1}\fi

\bibitem[{{Arentoft} {et~al.}(2007){Arentoft}, {de Ridder}, {Grundahl},
  {Glowienka}, {Waelkens}, {Dupret}, {Grigahc{\`e}ne}, {Lefever}, {Jensen},
  {Reyniers}, {Frandsen}, \& {Kjeldsen}}]{arentoft07}
{Arentoft}, T., {de Ridder}, J., {Grundahl}, F., {et~al.} 2007, \aap, 465, 965

\bibitem[{{Balona} {et~al.}(1994){Balona}, {Krisciunas}, \&
  {Cousins}}]{balona94}
{Balona}, L.~A., {Krisciunas}, K., \& {Cousins}, A.~W.~J. 1994, \mnras, 270,
  905

\bibitem[{{Bessell} {et~al.}(1998){Bessell}, {Castelli}, \& {Plez}}]{bessell98}
{Bessell}, M.~S., {Castelli}, F., \& {Plez}, B. 1998, \aap, 333, 231

\bibitem[{{Bikmaev} {et~al.}(2002){Bikmaev}, {Ryabchikova}, {Bruntt}, {Musaev},
  {Mashonkina}, {Belyakova}, {Shimansky}, {Barklem}, \&
  {Galazutdinov}}]{bikmaev02}
{Bikmaev}, I.~F., {Ryabchikova}, T.~A., {Bruntt}, H., {et~al.} 2002, \aap, 389,
  537

\bibitem[{{Boesgaard} \& {Tripicco}(1986)}]{boes86}
{Boesgaard}, A.~M. \& {Tripicco}, M.~J. 1986, \apj, 303, 724

\bibitem[{{Boesgaard} \& {Tripicco}(1987)}]{boes87}
{Boesgaard}, A.~M. \& {Tripicco}, M.~J. 1987, \apj, 313, 389

\bibitem[{{Breger} {et~al.}(2005){Breger}, {Lenz}, {Antoci}, {Guggenberger},
  {Shobbrook}, {Handler}, {Ngwato}, {Rodler}, {Rodriguez}, {L{\'o}pez de Coca},
  {Rolland}, \& {Costa}}]{breger05}
{Breger}, M., {Lenz}, P., {Antoci}, V., {et~al.} 2005, \aap, 435, 955

\bibitem[{{Bruntt} {et~al.}(2004){Bruntt}, {Bikmaev}, {Catala}, {Solano},
  {Gillon}, {Magain}, {Van't Veer-Menneret}, {St{\"u}tz}, {Weiss}, {Ballereau},
  {Bouret}, {Charpinet}, {Hua}, {Katz}, {Ligni{\`e}res}, \&
  {Lueftinger}}]{bruntt04}
{Bruntt}, H., {Bikmaev}, I.~F., {Catala}, C., {et~al.} 2004, \aap, 425, 683

\bibitem[{{Bruntt} {et~al.}(2002){Bruntt}, {Catala}, {Garrido},
  {Rodr{\'{\i}}guez}, {St{\"u}tz}, {Knoglinger}, {Mittermayer}, {Bouret},
  {Hua}, {Ligni{\`e}res}, {Charpinet}, {Van't Veer-Menneret}, \&
  {Ballereau}}]{bruntt02}
{Bruntt}, H., {Catala}, C., {Garrido}, R., {et~al.} 2002, \aap, 389, 345

\bibitem[{{Bruntt} {et~al.}(2007){Bruntt}, {Stello}, {Su{\'a}rez}, {Arentoft},
  {Bedding}, {Bouzid}, {Csubry}, {Dall}, {Dind}, {Frandsen}, {Gilliland},
  {Jacob}, {Jensen}, {Kang}, {Kim}, {Kiss}, {Kjeldsen}, {Koo}, {Lee}, {Lee},
  {Nuspl}, {Sterken}, \& {Szab{\'o}}}]{bruntt07}
{Bruntt}, H., {Stello}, D., {Su{\'a}rez}, J.~C., {et~al.} 2007, \mnras, 378,
  1371

\bibitem[{{Cameron}(1966)}]{cameron66}
{Cameron}, R.~C. 1966, Georgetown Observatory Monogram, 21, 0

\bibitem[{{Canuto} \& {Mazzitelli}(1992)}]{canuto92}
{Canuto}, V.~M. \& {Mazzitelli}, I. 1992, \apj, 389, 724

\bibitem[{{Cardelli} {et~al.}(1989){Cardelli}, {Clayton}, \&
  {Mathis}}]{cardelli89}
{Cardelli}, J.~A., {Clayton}, G.~C., \& {Mathis}, J.~S. 1989, \apj, 345, 245

\bibitem[{{Crawford} {et~al.}(1966){Crawford}, {Barnes}, {Faure}, \&
  {Golson}}]{crawford66}
{Crawford}, D.~L., {Barnes}, J.~V., {Faure}, B.~Q., \& {Golson}, J.~C. 1966,
  \aj, 71, 709

\bibitem[{{Cutri} {et~al.}(2003){Cutri}, {Skrutskie}, {van Dyk}, {Beichman},
  {Carpenter}, {Chester}, {Cambresy}, {Evans}, {Fowler}, {Gizis}, {Howard},
  {Huchra}, {Jarrett}, {Kopan}, {Kirkpatrick}, {Light}, {Marsh}, {McCallon},
  {Schneider}, {Stiening}, {Sykes}, {Weinberg}, {Wheaton}, {Wheelock}, \&
  {Zacarias}}]{cutri03}
{Cutri}, R.~M., {Skrutskie}, M.~F., {van Dyk}, S., {et~al.} 2003, {2MASS All
  Sky Catalog of point sources.} (The IRSA 2MASS All-Sky Point Source Catalog,
  NASA/IPAC Infrared Science
  Archive.~http://irsa.ipac.caltech.edu/applications/Gator/)

\bibitem[{{De Cat} {et~al.}(2006){De Cat}, {Eyer}, {Cuypers}, {Aerts},
  {Vandenbussche}, {Uytterhoeven}, {Reyniers}, {Kolenberg}, {Groenewegen},
  {Raskin}, {Maas}, \& {Jankov}}]{decat06}
{De Cat}, P., {Eyer}, L., {Cuypers}, J., {et~al.} 2006, \aap, 449, 281

\bibitem[{{Dupret} {et~al.}(2006){Dupret}, {Grigahc{\`e}ne}, {Garrido}, {De
  Ridder}, {Moya}, {Su{\'a}rez}, {Scuflaire}, {Gabriel}, \&
  {Goupil}}]{dupret06}
{Dupret}, M.-A., {Grigahc{\`e}ne}, A., {Garrido}, R., {et~al.} 2006, Memorie
  della Societa Astronomica Italiana, 77, 366

\bibitem[{{Dupret} {et~al.}(2004){Dupret}, {Grigahc{\`e}ne}, {Garrido},
  {Gabriel}, \& {Scuflaire}}]{dupret04}
{Dupret}, M.-A., {Grigahc{\`e}ne}, A., {Garrido}, R., {Gabriel}, M., \&
  {Scuflaire}, R. 2004, \aap, 414, L17

\bibitem[{{Dupret} {et~al.}(2005){Dupret}, {Grigahc{\`e}ne}, {Garrido},
  {Gabriel}, \& {Scuflaire}}]{dupretinstab}
{Dupret}, M.-A., {Grigahc{\`e}ne}, A., {Garrido}, R., {Gabriel}, M., \&
  {Scuflaire}, R. 2005, \aap, 435, 927

\bibitem[{{Erspamer} \& {North}(2003)}]{erspamer03}
{Erspamer}, D. \& {North}, P. 2003, \aap, 398, 1121

\bibitem[{{Gillon} \& {Magain}(2006)}]{gillon06}
{Gillon}, M. \& {Magain}, P. 2006, \aap, 448, 341

\bibitem[{{Gonzalez}(1998)}]{gonzalez98}
{Gonzalez}, G. 1998, \aap, 334, 221

\bibitem[{{Gray} \& {Kaye}(1999)}]{gray99}
{Gray}, R.~O. \& {Kaye}, A.~B. 1999, \aj, 118, 2993

\bibitem[{{Grevesse} {et~al.}(2007){Grevesse}, {Asplund}, \&
  {Sauval}}]{grevesse07}
{Grevesse}, N., {Asplund}, M., \& {Sauval}, A.~J. 2007, Space Science Reviews,
  105

\bibitem[{{Grevesse} \& {Sauval}(1998)}]{grevesse98}
{Grevesse}, N. \& {Sauval}, A.~J. 1998, Space Science Reviews, 85, 161

\bibitem[{{Handler}(1999)}]{handler99}
{Handler}, G. 1999, \mnras, 309, L19

\bibitem[{{Hauck} \& {Mermilliod}(1998)}]{hauck98}
{Hauck}, B. \& {Mermilliod}, M. 1998, \aaps, 129, 431

\bibitem[{{Heiter} {et~al.}(2002){Heiter}, {Kupka}, {van't Veer-Menneret},
  {Barban}, {Weiss}, {Goupil}, {Schmidt}, {Katz}, \& {Garrido}}]{heiter02}
{Heiter}, U., {Kupka}, F., {van't Veer-Menneret}, C., {et~al.} 2002, \aap, 392,
  619

\bibitem[{{Henry} \& {Fekel}(2005)}]{henry05}
{Henry}, G.~W. \& {Fekel}, F.~C. 2005, \aj, 129, 2026

\bibitem[{{Henry} {et~al.}(2007){Henry}, {Fekel}, \& {Henry}}]{henry07}
{Henry}, G.~W., {Fekel}, F.~C., \& {Henry}, S.~M. 2007, \aj, 133, 1421

\bibitem[{{Hinkle} {et~al.}(2000){Hinkle}, {Wallace}, {Valenti}, \&
  {Harmer}}]{hinkle}
{Hinkle}, K., {Wallace}, L., {Valenti}, J., \& {Harmer}, D. 2000, {Visible and
  Near Infrared Atlas of the Arcturus Spectrum 3727--9300~\AA} (San Francisco:
  ASP)

\bibitem[{{King} {et~al.}(2007){King}, {Matthews}, {Rowe}, {Cameron},
  {Kuschnig}, {Guenther}, {Moffat}, {Rucinski}, {Sasselov}, {Walker}, \&
  {Weiss}}]{king07}
{King}, H., {Matthews}, J.~M., {Rowe}, J.~F., {et~al.} 2007, ArXiv e-prints,
  706

\bibitem[{{Krisciunas} \& {Handler}(1995)}]{kris95}
{Krisciunas}, K. \& {Handler}, G. 1995, Informational Bulletin on Variable
  Stars, 4195, 1

\bibitem[{{Kupka}(1996)}]{kupkaconv}
{Kupka}, F. 1996, in ASP Conf. Ser. 108: M.A.S.S., Model Atmospheres and
  Spectrum Synthesis, ed. S.~J. {Adelman}, F.~{Kupka}, \& W.~W. {Weiss}, 73

\bibitem[{{Kupka} \& {Bruntt}(2001)}]{kupka01}
{Kupka}, F. \& {Bruntt}, H. 2001, {First COROT/MONS/MOST Ground Support
  Workshop, ed.\ C.~Sterken} (Vrije Universiteit Brussel, Belgium), 39

\bibitem[{{Kupka} {et~al.}(1999){Kupka}, {Piskunov}, {Ryabchikova}, {Stempels},
  \& {Weiss}}]{vald}
{Kupka}, F., {Piskunov}, N., {Ryabchikova}, T.~A., {Stempels}, H.~C., \&
  {Weiss}, W.~W. 1999, \aaps, 138, 119

\bibitem[{{Kurucz}(1993)}]{kurucz}
{Kurucz}, R. 1993, SYNTHE Spectrum Synthesis Programs and Line Data.~Kurucz
  CD-ROM No.~18.~Cambridge, Mass.: Smithsonian Astrophysical Observatory

\bibitem[{{Lejeune} \& {Schaerer}(2001)}]{lejeune01}
{Lejeune}, T. \& {Schaerer}, D. 2001, \aap, 366, 538

\bibitem[{{Masana} {et~al.}(2006){Masana}, {Jordi}, \& {Ribas}}]{masana06}
{Masana}, E., {Jordi}, C., \& {Ribas}, I. 2006, \aap, 450, 735

\bibitem[{{Mathias} {et~al.}(2004){Mathias}, {Le Contel}, {Chapellier},
  {Jankov}, {Sareyan}, {Poretti}, {Garrido}, {Rodr{\'{\i}}guez}, {Arellano
  Ferro}, {Alvarez}, {Parrao}, {Pe{\~n}a}, {Eyer}, {Aerts}, {De Cat}, {Weiss},
  \& {Zhou}}]{mathias04}
{Mathias}, P., {Le Contel}, J.-M., {Chapellier}, E., {et~al.} 2004, \aap, 417,
  189

\bibitem[{{McWilliam}(1998)}]{mcwilliam98}
{McWilliam}, A. 1998, \aj, 115, 1640

\bibitem[{{Moya} {et~al.}(2005){Moya}, {Su{\'a}rez}, {Amado},
  {Martin-Ru{\'{\i}}z}, \& {Garrido}}]{moya05}
{Moya}, A., {Su{\'a}rez}, J.~C., {Amado}, P.~J., {Martin-Ru{\'{\i}}z}, S., \&
  {Garrido}, R. 2005, \aap, 432, 189

\bibitem[{{Olsen}(1983)}]{olsen83}
{Olsen}, E.~H. 1983, \aaps, 54, 55

\bibitem[{{Olsen}(1988)}]{olsen88}
{Olsen}, E.~H. 1988, \aap, 189, 173

\bibitem[{{Paunzen} \& {Maitzen}(1998)}]{paunzen98lamboo}
{Paunzen}, E. \& {Maitzen}, H.~M. 1998, \aaps, 133, 1

\bibitem[{{Perry}(1991)}]{perry91}
{Perry}, C.~L. 1991, \pasp, 103, 494

\bibitem[{{Popper}(1980)}]{popper80}
{Popper}, D.~M. 1980, \araa, 18, 115

\bibitem[{{Renson} {et~al.}(1991){Renson}, {Gerbaldi}, \&
  {Catalano}}]{renson91}
{Renson}, P., {Gerbaldi}, M., \& {Catalano}, F.~A. 1991, \aaps, 89, 429

\bibitem[{{Rentzsch-Holm}(1996)}]{rent96}
{Rentzsch-Holm}, I. 1996, \aap, 312, 966

\bibitem[{{Rodr{\'{\i}}guez} {et~al.}(2006){Rodr{\'{\i}}guez}, {Costa}, {Zhou},
  {Grigahc{\`e}ne}, {Dupret}, {Su{\'a}rez}, {Moya}, {L{\'o}pez-Gonz{\'a}lez},
  {Wei}, \& {Fan}}]{rodriguez06}
{Rodr{\'{\i}}guez}, E., {Costa}, V., {Zhou}, A.-Y., {et~al.} 2006, \aap, 456,
  261

\bibitem[{{Rogers}(1995)}]{rogers95}
{Rogers}, N.~Y. 1995, Communications in Asteroseismology, 78, 1

\bibitem[{{Rowe} {et~al.}(2006){Rowe}, {Matthews}, {Cameron}, {Bohlender},
  {King}, {Kuschnig}, {Guenther}, {Moffat}, {Rucinski}, {Sasselov}, {Walker},
  \& {Weiss}}]{rowe06}
{Rowe}, J.~F., {Matthews}, J.~M., {Cameron}, C., {et~al.} 2006, Communications
  in Asteroseismology, 148, 34

\bibitem[{{Sadakane}(2006)}]{sadakane06}
{Sadakane}, K. 2006, \pasj, 58, 1023

\bibitem[{{Scardia} {et~al.}(2007){Scardia}, {Argyle}, {Prieur}, {Pansecchi},
  {Basso}, {Law}, \& {Mackay}}]{scardia07}
{Scardia}, M., {Argyle}, R.~W., {Prieur}, J.-L., {et~al.} 2007, Astronomische
  Nachrichten, 328, 146

\bibitem[{{Smalley}(2005)}]{smalley05}
{Smalley}, B. 2005, Memorie della Societa Astronomica Italiana Supplement, 8,
  130

\bibitem[{{Smalley} {et~al.}(2002){Smalley}, {Gardiner}, {Kupka}, \&
  {Bessell}}]{smalley02}
{Smalley}, B., {Gardiner}, R.~B., {Kupka}, F., \& {Bessell}, M.~S. 2002, \aap,
  395, 601

\bibitem[{{Stetson}(1991)}]{stetson91}
{Stetson}, P.~B. 1991, \aj, 102, 589

\bibitem[{{Su{\'a}rez} {et~al.}(2006{\natexlab{a}}){Su{\'a}rez}, {Garrido}, \&
  {Goupil}}]{suarez06b}
{Su{\'a}rez}, J.~C., {Garrido}, R., \& {Goupil}, M.~J. 2006{\natexlab{a}},
  \aap, 447, 649

\bibitem[{{Su{\'a}rez} {et~al.}(2006{\natexlab{b}}){Su{\'a}rez}, {Goupil}, \&
  {Morel}}]{suarez06a}
{Su{\'a}rez}, J.~C., {Goupil}, M.~J., \& {Morel}, P. 2006{\natexlab{b}}, \aap,
  449, 673

\bibitem[{{Valenti} \& {Piskunov}(1996)}]{sme}
{Valenti}, J.~A. \& {Piskunov}, N. 1996, \aaps, 118, 595

\bibitem[{{Zima} {et~al.}(2006){Zima}, {Wright}, {Bentley}, {Cottrell},
  {Heiter}, {Mathias}, {Poretti}, {Lehmann}, {Montemayor}, \&
  {Breger}}]{zima06}
{Zima}, W., {Wright}, D., {Bentley}, J., {et~al.} 2006, \aap, 455, 235

\end{thebibliography}



\appendix

\section{Detailed abundance results\label{sec:appab}}

In Tables~\ref{tab:ab}--\ref{tab:ab3}, we list the 
abundances of individual elements of the target stars. 
The abundances are measured differentially line-by-line
with respect to an observed spectrum of the Sun published 
by \cite{hinkle}.
The tables list the uncertainty on the mean value (in parenthesis) 
and the number of lines used in the analysis.
For example, the abundance of Carbon in HD~14940 is
measured to be $\log N_{\rm C}/N_{\rm total}-(\log N_{\rm C}/N_{\rm total})_\odot =-0.38\pm0.03$
from four lines. The quoted uncertainty is an internal value
and does not include contributions from the uncertainties 
on the adopted fundamental parameters or shortcomings of the applied model atmosphere,
which contribute by about $0.08$~dex on the uncertainty of the abundances (\cf\ Sect.~\ref{sec:uncer}).

\begin{table*}
 \centering
 \caption{Abundances for the bona fide and candidate \gamdor\ stars. The results are shown in the 
\topp\ \panel\ in Fig.~\ref{fig:abund}.
 \label{tab:ab}}
 \setlength{\tabcolsep}{3pt} 
 \begin{footnotesize}
\begin{tabular}{l|lr|lr|lr|lr|lr|lr|lr|lr}
          & \multicolumn{2}{c|}{    HD 14940} & \multicolumn{2}{c|}{    HD 40745} & \multicolumn{2}{c|}{    HD 48501} & \multicolumn{2}{c|}{ HD 110379} & \multicolumn{2}{c|}{ HD 126516} & \multicolumn{2}{c|}{   HD 135825} & \multicolumn{2}{c}{   HD 167858}  \\
\hline
  {C  \sc  i} &  $-0.38(3)$  &       4 & $-0.15    $ &       2 & $+0.04    $ &       1 & $-0.26(3)$  &       5 & $-0.17(4)$  &       3 & $-0.30    $ &       1 & $+0.08(2)$  &       7  \\ 
  {O  \sc  i} &  $-0.19    $ &       1 & $-0.30    $ &       1 & $+0.03    $ &       1 &         $-$ &     $-$ &         $-$ &     $-$ & $-0.01    $ &       1 & $-0.17    $ &       2  \\ 
  {Na \sc  i} &  $+0.00    $ &       2 &         $-$ &     $-$ &         $-$ &     $-$ & $-0.02    $ &       2 & $-0.21(3)$  &       5 & $+0.07    $ &       2 & $+0.23(4)$  &       3  \\ 
  {Mg \sc  i} &  $+0.20(7)$  &       5 &         $-$ &     $-$ &         $-$ &     $-$ & $+0.27(7)$  &       3 & $+0.07(7)$  &       4 & $+0.23(7)$  &       3 & $+0.35    $ &       2  \\ 
  {S  \sc  i} &  $+0.07    $ &       2 & $+0.20    $ &       1 & $+0.23    $ &       1 & $-0.15(3)$  &       3 &         $-$ &     $-$ & $+0.19    $ &       2 & $+0.14    $ &       2  \\ 
  {Si \sc  i} &  $-0.01(2)$  &      10 & $+0.09(2)$  &      16 & $+0.13(3)$  &       6 & $-0.02(2)$  &      15 & $-0.14(2)$  &       8 & $+0.02(2)$  &      12 & $+0.20(2)$  &      11  \\ 
  {Si \sc ii} &  $-0.19(6)$  &       3 & $+0.31(5)$  &       4 & $-0.04(5)$  &       3 & $+0.09    $ &       2 & $-0.06    $ &       1 & $+0.36(5)$  &       4 &         $-$ &     $-$  \\ 
  {Ca \sc  i} &  $+0.15(3)$  &      20 & $+0.16(4)$  &      14 & $+0.26(4)$  &       9 & $+0.06(3)$  &      17 & $-0.08(3)$  &      14 & $+0.17(3)$  &      20 & $+0.30(4)$  &       7  \\ 
  {Sc \sc ii} &  $+0.03(5)$  &       4 & $-0.05(5)$  &       4 & $+0.16(4)$  &       4 & $-0.21(5)$  &       7 & $-0.19(3)$  &       8 & $+0.18(4)$  &       6 & $+0.24(2)$  &       7  \\ 
  {Ti \sc  i} &  $+0.16(6)$  &       4 & $-0.05   $  &       2 &         $-$ &     $-$ & $-0.06    $ &       2 & $-0.07(3)$  &      14 & $-0.12(6)$  &       4 & $+0.11(4)$  &       5  \\ 
  {Ti \sc ii} &  $+0.04(3)$  &      10 & $+0.09(3)$  &      11 & $+0.20(6)$  &       6 & $-0.10(2)$  &      33 & $-0.17(2)$  &      12 & $+0.01(3)$  &      12 & $+0.18(3)$  &       7  \\ 
  {Cr \sc  i} &  $-0.01(4)$  &       9 & $-0.11(4)$  &       9 & $+0.08(6)$  &       3 & $-0.04(3)$  &      14 & $-0.24(4)$  &       9 & $+0.15(4)$  &      10 & $+0.26(3)$  &       9  \\ 
  {Cr \sc ii} &  $-0.05(3)$  &       7 & $-0.05(4)$  &       7 & $+0.27(5)$  &       5 & $-0.08(2)$  &      17 & $-0.28(2)$  &       9 & $+0.15(2)$  &      13 & $+0.19(2)$  &       9  \\ 
  {Mn \sc  i} &  $-0.59   $  &       2 & $-0.61   $  &       2 & $-0.47(8)$  &       4 & $-0.25(4)$  &       8 & $-0.46(4)$  &      10 & $-0.35(8)$  &       4 & $+0.03(3)$  &      11  \\ 
  {Mn \sc ii} &          $-$ &     $-$ &         $-$ &     $-$ &         $-$ &     $-$ &         $-$ &     $-$ &         $-$ &     $-$ &         $-$ &     $-$ &         $-$ &     $-$  \\ 
  {Fe \sc  i} &  $-0.06(1)$  &      42 & $-0.08(1)$  &      89 & $-0.08(2)$  &      20 & $+0.02(0)$  &     146 & $-0.21(1)$  &     133 & $+0.01(2)$  &      41 & $+0.27(1)$  &      83  \\ 
  {Fe \sc ii} &  $+0.00(2)$  &      13 & $+0.13(3)$  &      12 & $+0.11(3)$  &       8 & $-0.03(2)$  &      17 & $-0.25(2)$  &      19 & $+0.25(2)$  &      13 & $+0.23(2)$  &      12  \\ 
  {Co \sc  i} &          $-$ &     $-$ &         $-$ &     $-$ &         $-$ &     $-$ & $-0.05    $ &       1 & $-0.04    $ &       2 &         $-$ &     $-$ & $+0.24    $ &       1  \\ 
  {Ni \sc  i} &  $-0.00(4)$  &       8 & $-0.16(4)$  &       9 & $-0.23(4)$  &       7 & $-0.08(2)$  &      27 & $-0.20(1)$  &      29 & $-0.01(3)$  &      12 & $+0.26(2)$  &      16  \\ 
  {Cu \sc  i} &          $-$ &     $-$ &         $-$ &     $-$ &         $-$ &     $-$ & $-0.45    $ &       2 & $-0.55    $ &       2 &         $-$ &     $-$ & $-0.56    $ &       1  \\ 
  {Zn \sc  i} &          $-$ &     $-$ &         $-$ &     $-$ &         $-$ &     $-$ & $-0.26(7)$  &       3 & $-0.33(5)$  &       3 &         $-$ &     $-$ & $-0.01    $ &       2  \\ 
  {Y  \sc  i} &          $-$ &     $-$ &         $-$ &     $-$ &         $-$ &     $-$ &         $-$ &     $-$ &         $-$ &     $-$ &         $-$ &     $-$ &         $-$ &     $-$  \\ 
  {Y  \sc ii} &          $-$ &     $-$ &         $-$ &     $-$ &         $-$ &     $-$ & $-0.01(5)$  &       4 & $-0.11(4)$  &       7 & $+0.23    $ &       2 & $+0.30(3)$  &       4  \\ 
  {Zr \sc ii} &          $-$ &     $-$ &         $-$ &     $-$ &         $-$ &     $-$ &         $-$ &     $-$ & $+0.05(6)$  &       4 &         $-$ &     $-$ & $+0.64    $ &       1  \\ 
  {Ba \sc ii} &  $+0.56(7)$  &       4 & $+0.26    $ &       2 & $+0.37    $ &       2 & $+0.46(7)$  &       5 & $-0.05(10)$ &       3 & $+0.54(7)$  &       4 &         $-$ &     $-$  \\ 
\end{tabular}
\end{footnotesize}
\end{table*}

\begin{table*}
 \centering
 \caption{Abundances for the constant stars, the \bfss\ star HD~125081, 
and the tertiary reference stars (HD~32115, HD37594, and HD~49933). 
The results are shown in the \bott\ \panel\ in Fig.~\ref{fig:abund}.
 \label{tab:ab2}}
 \setlength{\tabcolsep}{3pt} 
 \begin{footnotesize}
\begin{tabular}{l|lr|lr|lr|lr|lr|lr|lr}
    &  \multicolumn{2}{c|}{  HD~7455} & \multicolumn{2}{c|}{    HD~22001} & \multicolumn{2}{c|}{   HD~32115} & \multicolumn{2}{c|}{    HD~33262} & \multicolumn{2}{c|}{   HD~37594} & \multicolumn{2}{c|}{   HD~49933} & \multicolumn{2}{c}{ HD~125081}  \\
\hline
  {C  \sc  i}  & $-0.45    $ &       2 & $-0.38(2)$  &       6 & $-0.15(2)$  &       9 & $-0.25(3)$  &       4 & $-0.24(3)$  &       4 & $-0.56(4)$  &       3 & $-0.41(3)$  &       6  \\ 
  {O  \sc  i}  &         $-$ &     $-$ & $-0.29    $ &       2 & $-0.15    $ &       2 &         $-$ &     $-$ & $-0.32    $ &       2 & $-0.53    $ &       2 & $-0.31    $ &       2  \\ 
  {Na \sc  i}  & $-0.42    $ &       2 & $-0.19(3)$  &       7 & $+0.03    $ &       2 & $-0.21(3)$  &       6 & $-0.39    $ &       2 & $-0.36    $ &       2 & $-0.31    $ &       2  \\ 
  {Mg \sc  i}  & $-0.16(7)$  &       3 & $-0.15(3)$  &       7 & $+0.53(9)$  &       3 & $+0.03(4)$  &       7 &         $-$ &     $-$ &         $-$ &     $-$ & $-0.31(5)$  &       4  \\ 
  {S  \sc  i}  &         $-$ &     $-$ & $-0.36(3)$  &       4 & $-0.06(3)$  &       3 & $-0.22(3)$  &       5 & $-0.18    $ &       1 & $-0.36    $ &       1 & $-0.14(3)$  &       3  \\ 
  {Si \sc  i}  & $-0.32(1)$  &      14 & $-0.13(2)$  &      12 & $+0.03(2)$  &      15 & $-0.06(1)$  &      18 & $-0.22(2)$  &       9 & $-0.37(2)$  &       8 & $-0.16(2)$  &      11  \\ 
  {Si \sc ii}  & $-0.31    $ &       2 &         $-$ &     $-$ & $+0.25    $ &       2 & $-0.11    $ &       2 &         $-$ &     $-$ &         $-$ &     $-$ &         $-$ &     $-$  \\ 
  {Ca \sc  i}  & $-0.37(3)$  &      15 & $-0.17(2)$  &      16 & $+0.14(3)$  &      15 & $+0.01(3)$  &      11 & $-0.17(4)$  &       8 & $-0.50(5)$  &       5 & $-0.14(4)$  &       8  \\ 
  {Sc \sc ii}  & $-0.28(4)$  &       4 & $-0.21(2)$  &      11 & $+0.15(4)$  &       6 & $+0.00(2)$  &      14 & $-0.32(4)$  &       4 & $-0.45(4)$  &       3 & $-0.24(2)$  &      12  \\ 
  {Ti \sc  i}  & $-0.31(2)$  &      16 & $-0.11(3)$  &       9 & $+0.06(5)$  &       5 & $-0.11(3)$  &       7 &         $-$ &     $-$ & $-0.52(6)$  &       4 & $-0.44    $ &       2  \\ 
  {Ti \sc ii}  & $-0.32(3)$  &      10 & $-0.28(2)$  &      19 & $+0.07(2)$  &      29 & $-0.08(3)$  &       5 & $-0.37(3)$  &       7 & $-0.41(3)$  &       4 & $-0.41(3)$  &       7  \\ 
  {Cr \sc  i}  & $-0.44(2)$  &      19 & $-0.20(2)$  &      14 & $+0.07(2)$  &      20 & $-0.07(2)$  &      13 & $-0.49(5)$  &       4 & $-0.63(7)$  &       3 & $-0.31(5)$  &       5  \\ 
  {Cr \sc ii}  & $-0.44(3)$  &       6 & $-0.28(2)$  &      15 & $+0.05(2)$  &      12 & $-0.31(2)$  &       7 & $-0.39(3)$  &       6 & $-0.43(4)$  &       3 & $-0.38(3)$  &       8  \\ 
  {Mn \sc  i}  & $-0.63(3)$  &      12 & $-0.43(2)$  &      19 & $-0.10(3)$  &      10 & $-0.26(2)$  &      22 & $-0.63    $ &       2 &         $-$ &     $-$ & $-0.44(2)$  &      15  \\ 
  {Fe \sc  i}  & $-0.38(0)$  &     226 & $-0.26(0)$  &     108 & $+0.07(0)$  &     189 & $-0.04(0)$  &     103 & $-0.31(1)$  &      82 & $-0.44(1)$  &      86 & $-0.28(1)$  &      98  \\ 
  {Fe \sc ii}  & $-0.38(2)$  &      16 & $-0.29(1)$  &      28 & $+0.04(1)$  &      32 & $-0.07(2)$  &      21 & $-0.32(2)$  &      17 & $-0.44(2)$  &      12 & $-0.26(1)$  &      22  \\ 
  {Co \sc  i}  & $-0.40(4)$  &       3 & $-0.41(4)$  &       5 &         $-$ &     $-$ & $-0.27(1)$  &      16 &         $-$ &     $-$ &         $-$ &     $-$ &         $-$ &     $-$  \\ 
  {Ni \sc  i}  & $-0.39(1)$  &      45 & $-0.29(1)$  &      29 & $-0.01(1)$  &      33 & $-0.19(1)$  &      32 & $-0.37(2)$  &      11 & $-0.48(2)$  &      14 & $-0.16(2)$  &      17  \\ 
  {Cu \sc  i}  & $-0.77    $ &       2 & $-0.87(6)$  &       4 & $-0.47    $ &       2 & $-0.60(6)$  &       4 &         $-$ &     $-$ &         $-$ &     $-$ & $-0.53(7)$  &       3  \\ 
  {Zn \sc  i}  & $-0.35    $ &       2 & $-0.49(6)$  &       3 & $-0.24    $ &       2 & $-0.32(6)$  &       3 &         $-$ &     $-$ &         $-$ &     $-$ & $-0.25(6)$  &       3  \\ 
  {Y  \sc ii}  &         $-$ &     $-$ & $-0.22(3)$  &       8 & $-0.15(5)$  &       3 & $-0.04(3)$  &       8 & $-0.34(4)$  &       3 &         $-$ &     $-$ & $+0.21(2)$  &       8  \\ 
  {Zr \sc ii}  &         $-$ &     $-$ & $-0.35    $ &       2 &         $-$ &     $-$ &         $-$ &     $-$ &         $-$ &     $-$ &         $-$ &     $-$ & $+0.23(6)$  &       4  \\ 
  {Ba \sc ii}  &         $-$ &     $-$ & $-0.16    $ &       2 & $+0.65(7)$  &       5 & $+0.37(7)$  &       4 &         $-$ &     $-$ &         $-$ &     $-$ & $+0.80(11)$ &       4  \\ 
\end{tabular}
\end{footnotesize}
\end{table*}

\begin{table*}
 \centering
 \caption{Abundances in the target stars with high \vsini.
 \label{tab:ab3}}
 \setlength{\tabcolsep}{3pt} 
 \begin{footnotesize}
\begin{tabular}{l|lr|lr|lr|lr|lr|lr|lr}
      & \multicolumn{2}{c|}{    HD~12901} & \multicolumn{2}{c|}{    HD~26298} & \multicolumn{2}{c|}{ HD~27290} & \multicolumn{2}{c|}{    HD~27604} & \multicolumn{2}{c|}{ HD~65526} & \multicolumn{2}{c|}{    HD~85964} & \multicolumn{2}{c}{   HD~218225}  \\
\hline
  {C  \sc  i} &  $+0.26(3)$  &       3 & $-0.47    $ &       2 & $+0.10(4)$  &       4 &         $-$ &     $-$ &         $-$ &     $-$ & $-0.05    $ &       2 &         $-$ &     $-$  \\ 
  {O  \sc  i} &          $-$ &     $-$ &         $-$ &     $-$ & $+0.14    $ &       1 &         $-$ &     $-$ &         $-$ &     $-$ &         $-$ &     $-$ &         $-$ &     $-$  \\ 
  {Na \sc  i} &  $-0.29    $ &       2 &         $-$ &     $-$ & $-0.88(7)$  &       3 &         $-$ &     $-$ &         $-$ &     $-$ &         $-$ &     $-$ &         $-$ &     $-$  \\ 
  {Mg \sc  i} &  $-0.07    $ &       2 & $+0.07    $ &       2 &         $-$ &     $-$ &         $-$ &     $-$ &         $-$ &     $-$ &         $-$ &     $-$ &         $-$ &     $-$  \\ 
  {S  \sc  i} &  $-0.13    $ &       2 & $-0.14    $ &       2 & $+0.19(3)$  &       3 &         $-$ &     $-$ & $-0.02    $ &       1 & $+0.47    $ &       2 &         $-$ &     $-$  \\ 
  {Si \sc  i} &  $-0.40(3)$  &       7 & $-0.03(3)$  &       4 & $+0.06(2)$  &      12 & $+0.54(6)$  &       3 & $-0.25    $ &       2 & $+0.07(3)$  &       6 &         $-$ &     $-$  \\ 
  {Si \sc ii} &  $+0.15    $ &       2 & $-0.00    $ &       2 & $+0.63    $ &       2 &         $-$ &     $-$ & $+0.15    $ &       2 &         $-$ &     $-$ &         $-$ &     $-$  \\ 
  {Ca \sc  i} &  $+0.09(4)$  &      10 & $+0.02(4)$  &      11 & $+0.29(3)$  &      15 & $+0.16(7)$  &       5 &         $-$ &     $-$ & $+0.34(5)$  &       8 & $+0.81(5)$  &       8  \\ 
  {Sc \sc ii} &          $-$ &     $-$ & $-0.43(7)$  &       3 &         $-$ &     $-$ &         $-$ &     $-$ &         $-$ &     $-$ & $+0.14(7)$  &       3 &         $-$ &     $-$  \\ 
  {Ti \sc  i} &          $-$ &     $-$ &         $-$ &     $-$ & $+0.19    $ &       2 &         $-$ &     $-$ &         $-$ &     $-$ &         $-$ &     $-$ &         $-$ &     $-$  \\ 
  {Ti \sc ii} &  $-0.49(5)$  &       4 & $-0.38(6)$  &       5 & $-0.38(3)$  &       9 & $+0.16    $ &       2 &         $-$ &     $-$ & $+0.01(4)$  &       8 & $+0.65(7)$  &       4  \\ 
  {Cr \sc  i} &  $-0.60(6)$  &       4 & $-0.27    $ &       2 & $-0.07(4)$  &       6 &         $-$ &     $-$ &         $-$ &     $-$ & $-0.24(6)$  &       4 & $-0.22(7)$  &       3  \\ 
  {Cr \sc ii} &  $-0.34    $ &       2 & $-0.38(4)$  &       5 & $+0.00(3)$  &      10 &         $-$ &     $-$ &         $-$ &     $-$ & $-0.01(4)$  &       7 &         $-$ &     $-$  \\ 
  {Mn \sc  i} &          $-$ &     $-$ &         $-$ &     $-$ & $-0.18(8)$  &       4 & $+0.19    $ &       2 &         $-$ &     $-$ &         $-$ &     $-$ &         $-$ &     $-$  \\ 
  {Mn \sc ii} &          $-$ &     $-$ &         $-$ &     $-$ &         $-$ &     $-$ &         $-$ &     $-$ &         $-$ &     $-$ &         $-$ &     $-$ &         $-$ &     $-$  \\ 
  {Fe \sc  i} &  $-0.24(1)$  &      75 & $-0.30(1)$  &      75 & $+0.12(1)$  &     117 & $+0.13(1)$  &      73 & $-0.33(2)$  &      41 & $+0.10(1)$  &      56 & $+0.29(1)$  &      61  \\ 
  {Fe \sc ii} &  $-0.24(4)$  &       8 & $-0.29(2)$  &      15 & $+0.32(2)$  &      17 & $+0.13(3)$  &       9 & $-0.19(4)$  &       5 & $+0.12(3)$  &      14 & $+0.62(3)$  &      10  \\ 
  {Ni \sc  i} &  $+0.13(5)$  &       5 & $-0.32(5)$  &       6 & $+0.12(4)$  &      10 &         $-$ &     $-$ & $-0.37(7)$  &       3 & $-0.03(5)$  &       6 &         $-$ &     $-$  \\ 
  {Zn \sc  i} &          $-$ &     $-$ &         $-$ &     $-$ &         $-$ &     $-$ &         $-$ &     $-$ &         $-$ &     $-$ &         $-$ &     $-$ &         $-$ &     $-$  \\ 
  {Y  \sc  i} &          $-$ &     $-$ &         $-$ &     $-$ &         $-$ &     $-$ &         $-$ &     $-$ &         $-$ &     $-$ &         $-$ &     $-$ &         $-$ &     $-$  \\ 
  {Zr \sc ii} &          $-$ &     $-$ &         $-$ &     $-$ &         $-$ &     $-$ &         $-$ &     $-$ &         $-$ &     $-$ &         $-$ &     $-$ &         $-$ &     $-$  \\ 
  {Ba \sc ii} &  $+0.35    $ &       2 &         $-$ &     $-$ & $+0.56    $ &       2 &         $-$ &     $-$ &         $-$ &     $-$ &         $-$ &     $-$ &         $-$ &     $-$  \\ 
\end{tabular}
\end{footnotesize}
\end{table*}


\section{Line list for HD~110379\label{sec:app110}}

In Table~\ref{tab:linelist} we list 
the lines used in the abundance analysis of HD~110379 with
atomic parameters extracted from the \vald\ database \citep{vald}.
It represents a typical example of lines used in the analysis 
for the slowly rotating stars in our sample.


   \begin{table*}
      \caption[]{The atomic number, element name, 
wavelength, and \loggf\ from the \vald\ database 
for the lines used in the analysis of HD~110379.
         \label{tab:linelist}}
\centering                          
\begin{tiny} 
\begin{tabular}{r@{\hskip 0.25cm}c@{\hskip 0.25cm}c|r@{\hskip 0.25cm}c@{\hskip 0.25cm}c|r@{\hskip 0.25cm}c@{\hskip 0.25cm}c|r@{\hskip 0.25cm}c@{\hskip 0.25cm}c|r@{\hskip 0.25cm}c@{\hskip 0.25cm}c}  
\hline\hline

\eeee & $\lambda$ [\AA] & \loggf & \eeee & $\lambda$ [\AA] & \loggf & \eeee & $\lambda$ [\AA] & \loggf & \eeee & $\lambda$ [\AA] & \loggf & \eeee & $\lambda$ [\AA] & \loggf \\ 

\hline

 $^{ 6}$C\ione & 4770.026 & $-2.439$ & $$     Ti\ione & 5210.385 & $-0.884$ & $$     Fe\ione & 4757.582 & $-2.321$ & $$     Fe\ione & 5434.524 & $-2.122$ & $$ Fe\ione & 6408.018 & $-1.018$ \\
 $$     C\ione & 4932.049 & $-1.884$ & $$     Ti\itwo & 4501.273 & $-0.760$ & $$     Fe\ione & 4791.246 & $-2.435$ & $$     Fe\ione & 5441.339 & $-1.730$ & $$ Fe\ione & 6419.950 & $-0.240$ \\
 $$     C\ione & 5380.337 & $-1.842$ & $$     Ti\itwo & 4518.327 & $-2.640$ & $$     Fe\ione & 4802.880 & $-1.514$ & $$     Fe\ione & 5445.042 & $-0.020$ & $$ Fe\ione & 6421.351 & $-2.027$ \\
 $$     C\ione & 6587.610 & $-1.596$ & $$     Ti\itwo & 4544.028 & $-2.530$ & $$     Fe\ione & 4843.144 & $-1.840$ & $$     Fe\ione & 5466.396 & $-0.630$ & $$ Fe\ione & 6609.110 & $-2.692$ \\
\cline{1-3}
 $^{ 8}$O\ione & 6158.186 & $-0.409$ & $$     Ti\itwo & 4563.761 & $-0.790$ & $$     Fe\ione & 4909.384 & $-1.273$ & $$     Fe\ione & 5472.709 & $-1.495$ & $$ Fe\ione & 6677.987 & $-1.418$ \\
\cline{1-3}
$^{11}$Na\ione & 5688.205 & $-0.450$ & $$     Ti\itwo & 4589.958 & $-1.620$ & $$     Fe\ione & 4942.459 & $-1.409$ & $$     Fe\ione & 5473.900 & $-0.760$ & $$ Fe\ione & 6750.153 & $-2.621$ \\
$^{12}$Mg\ione & 4702.991 & $-0.666$ & $$     Ti\itwo & 4779.985 & $-1.260$ & $$     Fe\ione & 4946.388 & $-1.170$ & $$     Fe\ione & 5483.099 & $-1.407$ & $$ Fe\ione & 6810.263 & $-0.986$ \\
$$     Mg\ione & 5172.684 & $-0.402$ & $$     Ti\itwo & 4805.085 & $-0.960$ & $$     Fe\ione & 4966.089 & $-0.871$ & $$     Fe\ione & 5497.516 & $-2.849$ & $$ Fe\itwo & 4520.224 & $-2.600$ \\
$$     Mg\ione & 5183.604 & $-0.180$ & $$     Ti\itwo & 5010.212 & $-1.300$ & $$     Fe\ione & 4967.890 & $-0.622$ & $$     Fe\ione & 5501.465 & $-3.047$ & $$ Fe\itwo & 4541.524 & $-2.790$ \\
$$     Mg\ione & 5528.405 & $-0.620$ & $$     Ti\itwo & 5013.677 & $-1.990$ & $$     Fe\ione & 4969.918 & $-0.710$ & $$     Fe\ione & 5506.779 & $-2.797$ & $$ Fe\itwo & 4576.340 & $-2.920$ \\
$$     Mg\ione & 5711.088 & $-1.833$ & $$     Ti\itwo & 5129.152 & $-1.300$ & $$     Fe\ione & 4973.102 & $-0.950$ & $$     Fe\ione & 5543.150 & $-1.570$ & $$ Fe\itwo & 4620.521 & $-3.240$ \\
\cline{1-3}
$^{13}$Al\ione & 6696.023 & $-1.347$ & $$     Ti\itwo & 5211.536 & $-1.356$ & $$     Fe\ione & 4988.950 & $-0.890$ & $$     Fe\ione & 5560.212 & $-1.190$ & $$ Fe\itwo & 4731.453 & $-3.000$ \\
$$     Al\ione & 6698.673 & $-1.647$ & $$     Ti\itwo & 5381.015 & $-1.970$ & $$     Fe\ione & 4994.130 & $-3.080$ & $$     Fe\ione & 5563.600 & $-0.990$ & $$ Fe\itwo & 5120.352 & $-4.214$ \\
\cline{1-3}
$^{14}$Si\ione & 5645.613 & $-2.140$ & $$     Ti\itwo & 5490.690 & $-2.650$ & $$     Fe\ione & 5014.943 & $-0.303$ & $$     Fe\ione & 5569.618 & $-0.486$ & $$ Fe\itwo & 5256.938 & $-4.250$ \\
$$     Si\ione & 5675.417 & $-1.030$ & $$     Ti\itwo & 6491.561 & $-1.793$ & $$     Fe\ione & 5027.120 & $-0.559$ & $$     Fe\ione & 5576.089 & $-1.000$ & $$ Fe\itwo & 5362.869 & $-2.739$ \\
\cline{4-6}
$$     Si\ione & 5708.400 & $-1.470$ & $^{24}$Cr\ione & 4626.174 & $-1.320$ & $$     Fe\ione & 5028.126 & $-1.123$ & $$     Fe\ione & 5586.756 & $-0.120$ & $$ Fe\itwo & 6084.111 & $-3.780$ \\
$$     Si\ione & 5747.667 & $-0.780$ & $$     Cr\ione & 4646.148 & $-0.700$ & $$     Fe\ione & 5029.618 & $-2.050$ & $$     Fe\ione & 5633.947 & $-0.270$ & $$ Fe\itwo & 6147.741 & $-2.721$ \\
$$     Si\ione & 5753.623 & $-0.830$ & $$     Cr\ione & 4718.426 & $+ 0.090$ & $$     Fe\ione & 5054.643 & $-1.921$ & $$     Fe\ione & 5638.262 & $-0.870$ & $$  Fe\itwo & 6149.258 & $-2.720$ \\
$$     Si\ione & 6125.021 & $-0.930$ & $$     Cr\ione & 5204.506 & $-0.208$ & $$     Fe\ione & 5067.150 & $-0.970$ & $$     Fe\ione & 5686.530 & $-0.446$ & $$ Fe\itwo & 6238.392 & $-2.630$ \\
$$     Si\ione & 6131.852 & $-1.140$ & $$     Cr\ione & 5206.038 & $+ 0.019$ & $$     Fe\ione & 5074.748 & $-0.200$ & $$     Fe\ione & 5701.545 & $-2.216$ & $$  Fe\itwo & 6247.557 & $-2.310$ \\
$$     Si\ione & 6145.016 & $-0.820$ & $$     Cr\ione & 5208.419 & $+ 0.158$ & $$     Fe\ione & 5076.262 & $-0.767$ & $$     Fe\ione & 5705.992 & $-0.530$ & $$  Fe\itwo & 6416.919 & $-2.650$ \\
$$     Si\ione & 6194.416 & $-1.900$ & $$     Cr\ione & 5296.691 & $-1.400$ & $$     Fe\ione & 5090.774 & $-0.400$ & $$     Fe\ione & 5717.833 & $-1.130$ & $$ Fe\itwo & 6432.680 & $-3.520$ \\
\cline{13-15}
$$     Si\ione & 6237.319 & $-0.530$ & $$     Cr\ione & 5297.376 & $+ 0.167$ & $$     Fe\ione & 5121.639 & $-0.810$ & $$     Fe\ione & 5731.762 & $-1.300$ & $^{27}$Co\ione & 5342.695 & $+ 0.690$ \\
\cline{13-15}
$$     Si\ione & 6243.815 & $-0.770$ & $$     Cr\ione & 5348.312 & $-1.290$ & $$     Fe\ione & 5123.720 & $-3.068$ & $$     Fe\ione & 5752.023 & $-1.267$ & $^{28}$Ni\ione & 4715.757 & $-0.320$ \\
$$     Si\ione & 6244.466 & $-0.690$ & $$     Cr\ione & 5787.965 & $-0.083$ & $$     Fe\ione & 5131.469 & $-2.515$ & $$     Fe\ione & 5762.992 & $-0.450$ & $$ Ni\ione & 4756.510 & $-0.270$ \\
$$     Si\ione & 6254.188 & $-0.600$ & $$     Cr\itwo & 4554.988 & $-1.282$ & $$     Fe\ione & 5133.689 & $+ 0.140$ & $$     Fe\ione & 5809.218 & $-1.840$ & $$  Ni\ione & 4829.016 & $-0.330$ \\
$$     Si\ione & 6414.980 & $-1.100$ & $$     Cr\itwo & 4558.650 & $-0.449$ & $$     Fe\ione & 5141.739 & $-1.964$ & $$     Fe\ione & 5816.373 & $-0.601$ & $$ Ni\ione & 4904.407 & $-0.170$ \\
$$     Si\itwo & 6347.109 & $+ 0.297$ & $$     Cr\itwo & 4588.199 & $-0.627$ & $$     Fe\ione & 5150.840 & $-3.003$ & $$     Fe\ione & 5859.578 & $-0.398$ & $$  Ni\ione & 4935.831 & $-0.350$ \\
$$     Si\itwo & 6371.371 & $-0.003$ & $$     Cr\itwo & 4634.070 & $-0.990$ & $$     Fe\ione & 5151.911 & $-3.322$ & $$     Fe\ione & 5862.353 & $-0.058$ & $$ Ni\ione & 4937.341 & $-0.390$ \\
\cline{1-3}
 $^{16}$S\ione & 6046.027 & $-1.030$ & $$     Cr\itwo & 4824.127 & $-0.970$ & $$     Fe\ione & 5159.058 & $-0.820$ & $$     Fe\ione & 5883.817 & $-1.360$ & $$ Ni\ione & 4980.166 & $+ 0.070$ \\
 $$     S\ione & 6052.674 & $-0.740$ & $$     Cr\itwo & 5237.329 & $-1.160$ & $$     Fe\ione & 5162.273 & $+ 0.020$ & $$     Fe\ione & 5930.180 & $-0.230$ & $$  Ni\ione & 4998.218 & $-0.690$ \\
 $$     S\ione & 6757.171 & $-0.310$ & $$     Cr\itwo & 5274.964 & $-1.290$ & $$     Fe\ione & 5194.942 & $-2.090$ & $$     Fe\ione & 5934.655 & $-1.170$ & $$ Ni\ione & 5081.107 & $+ 0.300$ \\
\cline{1-3}
$^{20}$Ca\ione & 4878.126 & $+ 0.430$ & $$     Cr\itwo & 5305.853 & $-2.357$ & $$     Fe\ione & 5198.711 & $-2.135$ & $$     Fe\ione & 5987.066 & $-0.556$ & $$  Ni\ione & 5082.339 & $-0.540$ \\
$$     Ca\ione & 5349.465 & $-1.178$ & $$     Cr\itwo & 5308.408 & $-1.846$ & $$     Fe\ione & 5217.389 & $-1.070$ & $$     Fe\ione & 6003.012 & $-1.120$ & $$ Ni\ione & 5084.089 & $+ 0.030$ \\
$$     Ca\ione & 5581.965 & $-0.569$ & $$     Cr\itwo & 5310.687 & $-2.280$ & $$     Fe\ione & 5228.377 & $-1.290$ & $$     Fe\ione & 6008.554 & $-1.078$ & $$ Ni\ione & 5155.762 & $+ 0.011$ \\
$$     Ca\ione & 5588.749 & $+ 0.313$ & $$     Cr\itwo & 5313.563 & $-1.650$ & $$     Fe\ione & 5242.491 & $-0.967$ & $$     Fe\ione & 6020.169 & $-0.270$ & $$  Ni\ione & 5663.975 & $-0.430$ \\
$$     Ca\ione & 5590.114 & $-0.596$ & $$     Cr\itwo & 5334.869 & $-1.562$ & $$     Fe\ione & 5243.777 & $-1.150$ & $$     Fe\ione & 6024.058 & $-0.120$ & $$ Ni\ione & 5694.977 & $-0.610$ \\
$$     Ca\ione & 5594.462 & $+ 0.051$ & $$     Cr\itwo & 5508.606 & $-2.110$ & $$     Fe\ione & 5250.646 & $-2.181$ & $$     Fe\ione & 6027.051 & $-1.089$ & $$  Ni\ione & 5715.066 & $-0.352$ \\
\cline{4-6}
$$     Ca\ione & 5598.480 & $-0.134$ & $^{25}$Mn\ione & 4754.042 & $-0.086$ & $$     Fe\ione & 5253.462 & $-1.573$ & $$     Fe\ione & 6056.005 & $-0.460$ & $$ Ni\ione & 5760.828 & $-0.800$ \\
$$     Ca\ione & 5857.451 & $+ 0.257$ & $$     Mn\ione & 4761.512 & $-0.138$ & $$     Fe\ione & 5281.790 & $-0.834$ & $$     Fe\ione & 6065.482 & $-1.530$ & $$  Ni\ione & 5805.213 & $-0.640$ \\
$$     Ca\ione & 6122.217 & $-0.386$ & $$     Mn\ione & 4762.367 & $+ 0.425$ & $$     Fe\ione & 5302.302 & $-0.720$ & $$     Fe\ione & 6078.491 & $-0.424$ & $$  Ni\ione & 6086.276 & $-0.530$ \\
$$     Ca\ione & 6162.173 & $-0.167$ & $$     Mn\ione & 4783.427 & $+ 0.042$ & $$     Fe\ione & 5315.070 & $-1.550$ & $$     Fe\ione & 6127.907 & $-1.399$ & $$  Ni\ione & 6116.174 & $-0.677$ \\
$$     Ca\ione & 6163.755 & $-1.303$ & $$     Mn\ione & 4823.524 & $+ 0.144$ & $$     Fe\ione & 5339.929 & $-0.647$ & $$     Fe\ione & 6136.615 & $-1.400$ & $$  Ni\ione & 6176.807 & $-0.260$ \\
$$     Ca\ione & 6166.439 & $-1.156$ & $$     Mn\ione & 5377.637 & $-0.109$ & $$     Fe\ione & 5341.024 & $-1.953$ & $$     Fe\ione & 6170.507 & $-0.440$ & $$ Ni\ione & 6767.768 & $-2.170$ \\
\cline{13-15}
$$     Ca\ione & 6169.042 & $-0.804$ & $$     Mn\ione & 6021.819 & $+ 0.034$ & $$     Fe\ione & 5361.625 & $-1.430$ & $$     Fe\ione & 6173.336 & $-2.880$ & $^{29}$Cu\ione & 5105.537 & $-1.516$ \\
\cline{4-6}
$$     Ca\ione & 6169.563 & $-0.527$ & $^{26}$Fe\ione & 4547.847 & $-1.012$ & $$     Fe\ione & 5364.871 & $+ 0.228$ & $$     Fe\ione & 6213.430 & $-2.482$ & $$  Cu\ione & 5782.127 & $-1.720$ \\
\cline{13-15}
$$     Ca\ione & 6439.075 & $+ 0.394$ & $$     Fe\ione & 4566.989 & $-2.080$ & $$     Fe\ione & 5373.709 & $-0.860$ & $$     Fe\ione & 6219.281 & $-2.433$ & $^{30}$Zn\ione & 4680.134 & $-0.815$ \\
$$     Ca\ione & 6449.808 & $-1.015$ & $$     Fe\ione & 4602.941 & $-2.209$ & $$     Fe\ione & 5379.574 & $-1.514$ & $$     Fe\ione & 6230.723 & $-1.281$ & $$ Zn\ione & 4722.153 & $-0.338$ \\
$$     Ca\ione & 6493.781 & $+ 0.019$ & $$     Fe\ione & 4607.647 & $-1.545$ & $$     Fe\ione & 5389.479 & $-0.410$ & $$     Fe\ione & 6232.641 & $-1.223$ & $$  Zn\ione & 4810.528 & $-0.137$ \\
\cline{13-15}
$$     Ca\ione & 6499.650 & $-0.719$ & $$     Fe\ione & 4613.203 & $-1.670$ & $$     Fe\ione & 5391.461 & $-0.825$ & $$     Fe\ione & 6252.555 & $-1.687$ &  $^{39}$Y\itwo & 4900.120 & $-0.090$ \\
$$     Ca\ione & 6717.681 & $-0.596$ & $$     Fe\ione & 4625.045 & $-1.340$ & $$     Fe\ione & 5393.168 & $-0.715$ & $$     Fe\ione & 6256.361 & $-2.408$ &  $$  Y\itwo & 5087.416 & $-0.170$ \\
\cline{1-3}
$^{21}$Sc\itwo & 4670.407 & $-0.576$ & $$     Fe\ione & 4632.912 & $-2.913$ & $$     Fe\ione & 5398.279 & $-0.670$ & $$     Fe\ione & 6265.134 & $-2.550$ &  $$  Y\itwo & 5200.406 & $-0.570$ \\
\cline{13-15}
$$     Sc\itwo & 5031.021 & $-0.400$ & $$     Fe\ione & 4638.010 & $-1.119$ & $$     Fe\ione & 5400.502 & $-0.160$ & $$     Fe\ione & 6270.225 & $-2.464$ & $^{40}$Zr\itwo & 5112.297 & $-0.590$ \\
\cline{13-15}
$$     Sc\itwo & 5239.813 & $-0.765$ & $$     Fe\ione & 4733.592 & $-2.988$ & $$     Fe\ione & 5405.775 & $-1.844$ & $$     Fe\ione & 6335.331 & $-2.177$ & $^{56}$Ba\itwo & 4934.076 & $-0.150$ \\
$$     Sc\itwo & 5684.202 & $-1.074$ & $$     Fe\ione & 4735.844 & $-1.325$ & $$     Fe\ione & 5410.910 & $+ 0.398$ & $$     Fe\ione & 6336.824 & $-0.856$ & $$  Ba\itwo & 5853.668 & $-1.000$ \\
$$     Sc\itwo & 6604.601 & $-1.309$ & $$     Fe\ione & 4736.773 & $-0.752$ & $$     Fe\ione & 5415.199 & $+ 0.642$ & $$     Fe\ione & 6338.877 & $-1.060$ & $$  Ba\itwo & 6141.713 & $-0.076$ \\
\cline{1-3}
$^{22}$Ti\ione & 4981.731 & $+ 0.504$ & $$     Fe\ione & 4745.800 & $-1.270$ & $$     Fe\ione & 5424.068 & $+ 0.520$ & $$     Fe\ione & 6380.743 & $-1.376$ & $$   Ba\itwo & 6496.897 & $-0.377$ \\

\end{tabular}
\end{tiny}
\end{table*}

\end{document}